\documentclass[prx,twocolumn,english,superscriptaddress,floatfix,longbibliography,amsmath]{revtex4-2}
\usepackage{hyperref}
\usepackage{graphicx}
\usepackage[graphicx]{realboxes}
\usepackage{babel}
\usepackage{times}
\usepackage{dcolumn}
\usepackage[usenames,dvipsnames]{color}
\usepackage{units}
\usepackage{bm}
\usepackage{soul}
\usepackage[utf8]{inputenc}
\usepackage{float}
\usepackage{dsfont} %for the indentity operator: \mathds{1}
\usepackage{xcolor}
\usepackage{rotating}

\begin{document}

\title{Generalized transmon Hamiltonian for Andreev spin qubits}

\author{Luka Pave\v{s}i\'{c}}
\email{luka.pavesic@ijs.si}
\affiliation{Jo\v{z}ef Stefan Institute, Jamova 39, SI-1000 Ljubljana, Slovenia}
\affiliation{Faculty of Mathematics and Physics, University of Ljubljana, Jadranska 19, SI-1000 Ljubljana, Slovenia}

\author{Rok \v{Z}itko}
\email{rok.zitko@ijs.si}
\affiliation{Jo\v{z}ef Stefan Institute, Jamova 39, SI-1000 Ljubljana, Slovenia}
\affiliation{Faculty of Mathematics and Physics, University of Ljubljana, Jadranska 19, SI-1000 Ljubljana, Slovenia}

\begin{abstract}

We solve the problem of an interacting quantum dot embedded in a Josephson junction between two superconductors with finite charging energy described by the transmon (Cooper pair box) Hamiltonian. 
The approach is based on the flat-band approximation of the Richardson model, which reduces the Hilbert space to the point where exact diagonalisation is possible while retaining all states that are necessary to describe the low energy phenomena. 
The presented method accounts for the physics of the quantum dot, the Josephson effect and the Coulomb repulsion (charging energy) at the same level. 
In particular, it captures the quantum fluctuations of the superconducting phase as well as the coupling between the superconducting phase and the quantum dot (spin) degrees of freedom.
The method can be directly applied for modelling Andreev spin qubits embedded in transmon circuits in all parameter regimes, for describing time-dependent processes, and for the calculation of transition matrix elements for microwave-driven transmon, spin-flip and mixed transitions that involve coupling to charge or current degree of freedom.
\end{abstract}

\maketitle

\newcommand{\REF}[1]{\textcolor{red}{REF(#1)}}
\newcommand{\red}[1]{\textcolor{red}{#1}}
\newcommand{\rz}[1]{\textcolor{teal}{#1}}

\newcommand{\ket}[1]{\vert #1 \rangle}
\newcommand{\bra}[1]{\langle #1 \vert}

\newcommand{\QD}{\mathrm{QD}}
\newcommand{\SC}{\mathrm{SC}}

\newcommand{\Ec}{E_c}
\newcommand{\EJ}{E_J}
\newcommand{\EJeff}{ E_J^\mathrm{eff} }
\newcommand{\EJref}{ E_J^\mathrm{ref} }
\newcommand{\phiext}{\phi_\mathrm{ext}}
\newcommand{\dm}{m}
\newcommand{\mtot}{m_\mathrm{tot}}

\newcommand{\ml}{m_L}
\newcommand{\mr}{m_R}

\newcommand{\id}{\mathds{1}}

\section{Introduction}
Superconducting circuits are one of the leading platforms for the realization of various quantum technological applications.
Most implementations of superconducting qubits are based on creating an anharmonic oscillator by replacing the inductor in an LC-circuit with a Josephson junction (JJ), which has non-linear inductance~\cite{Kjaergaard2020, Devoret_review, Devoret_lesHouches}.
The realization where the charging energy due to capacitance is small
compared to the Josephson energy is called the transmon qubit~\cite{Koch2007}. 
As all other superconducting devices, the transmons utilize the macroscopic coherence of superconducting states to encode and manipulate quantum information~\cite{Wendin2017}. Their popularity is due to their robustness with respect to charge fluctuations, which is one of the main decoherence mechanisms in superconducting qubits.

In pursuit of further enhancing these devices, a novel approach has emerged, combining the robust coherence of superconductors (SCs) with the controllability of spin qubits built out of semiconducting quantum dots (QD).
The idea consists of embedding a QD into the JJ and storing quantum information in the spin of the quasiparticle trapped in discrete subgap states that emerge in the few-channel regime of the JJ. The architecture is called the Andreev spin qubit (ASQ)~\cite{Hays2021, Pita-Vidal:2022, pitavidal2023strong, Chtchelkatchev:2003, Zazunov2003}. The spin-orbit coupling (SOC) permits manipulation of the spin degree of freedom using the supercurrent or the electric field, as well as advanced readout based on circuit quantum electrodynamics (cQED) techniques.
If the ASQ is embedded in a transmon, such setup can actually support two physical qubits, one defined in the QD spin and the other in the standard computational subspace of transmon excitations~\cite{Pita-Vidal:2022}. 

Modelling of the transmon qubit typically relies on neglecting the superconducting quasiparticles and only accounting for the dynamics of Cooper pairs~\cite{Koch2007, Devoret_review}. 
The SC gap---the energy scale for creating quasiparticles---is typically much larger than the energy of transmon excitations, so this is a good approximation.
However, the presence of an interacting QD in the JJ induces Cooper pair-breaking processes, and an accurate description of the physics requires solving the full electronic problem. 
This is particularly important for modelling the ASQ, where a QD with a large charging energy is favored as it ensures a ground state with a single localized spin trapped in the QD. 

Two non-interacting SC leads coupled to an interacting QD constitute a quantum impurity problem, which is numerically solvable with standard impurity solvers, such as the numerical renormalization group~\cite{Wilson1975, Bulla2008}. The numerical procedure hinges on the separation of high and low energy scales, a feature of non-interacting leads.
However, simultaneously including also the charging energy of the SC islands---a critical feature of the transmon---makes the leads interacting. 
As far as we are aware, a method for attacking such problems does not exist.

In this work, we present a method for solving the microscopic QD-transmon model.
It is based on the flat-band approximation for the SC islands, which are described by a single active orbital coupled to a condensate of Cooper pairs.
This approach exponentially reduces the Hilbert space to the point where exact numerical diagonalisation is possible for a system of thousands of electrons, while retaining the key subspace  for capturing the low-energy phenomena. 

The paper is formatted as follows:
In Section~\ref{sec:known_properties} we describe the basic properties of the QD--transmon system we aim to reproduce. 
In Sec.~\ref{sec:model} we introduce the model and the reduced basis.
Section~\ref{sec:results} contains benchmark results that validate the method. We reproduce known behaviour and point out parameter regimes where the standard transmon approximation is not adequate. 
Section~\ref{sec:outlook} presents three specific examples of possible applications of our approach to problems that cannot be solved in any other way: 1) the case where the QD and SC degrees of freedom are strongly coupled, 2) the case of time-dependent perturbations, such as a microwave pulse, 3) the calculation of transition matrix elements involving both qubit degrees of freedom.

\section{Basic properties of the quantum dot Josephson junctions}
\label{sec:known_properties}

In this section we review the basic properties of quantum dot Josephson junctions to set the stage for the discussion of the full model.
The focus is on the interplay of electronic processes that determine the nature of low-energy states in the superconducting gap. 

\subsection{Cooper pair tunneling}

\newcommand{\mm}{m}

In a conventional JJ without an embedded QD quasiparticles play no role at temperatures much lower than the superconducting gap $\Delta$. An approximate description in terms of Cooper pair hopping is adequate to capture the Josephson effect. 
Microscopically, it is shown to arise form the coherent transfer of electron pairs across the junction~\cite{tinkham, Gobert_2004, Glazman2021}.

This can be expressed by writing the Hamiltonian in the charge basis $\ket{\mm}$, with $\mm = 0, 1, -1, \dots$ the \textit{difference} in the number of Cooper pairs occupying the two SCs.
We allow processes where a single Cooper pair hops across the junction. These terms couple $\ket{\mm}$ to the neighbouring $\ket{\mm \pm 1}$, and the Hamiltonian in this basis is  tridiagonal~\cite{Gobert_2004}.
As the different $\ket{\mm}$ states are equivalent, this corresponds to a tight-binding chain in $\mm$-space.
It is diagonalised by Fourier transforming $\ket{\mm}$ into states labelled by the dual quantity---this is the emergent superconducting phase difference $\phi$.
The eigenstates are Bloch waves, superpositions of $\ket{m}$ states, $ \ket{\phi} \propto \sum_\mm e^{i \mm \phi}\ket{\mm}$.
Their dispersion is given by
\begin{equation}
    E(\phi) = -\EJ \cos \phi,
\end{equation}
where the emergent energy scale, the Josephson energy $\EJ$, is given by twice the hopping matrix element between $\ket{\mm}$ and $\ket{\mm\pm1}$, i.e., the energy associated with transferring a single Cooper pair across the junction.

\subsection{Charging energy}

In nanoscopic structures, the small size of the SCs that form the junction implies large Coulomb repulsion between the electrons occupying the device \cite{Tuominen1992,Lafarge1993}.
This is encapsulated in the charging energy $\Ec$, which also includes other capacitance effects (such as the capacitive shunting of the JJ  in the transmon~\cite{Koch2007, Kjaergaard2020}). 

The problem is described by the transmon Hamiltonian (also known as the Cooper pair box Hamiltonian), which combines pair hopping and charging terms.
It was first introduced to describe the Cooper pair box~\cite{Bouchiat1998, Nakamura1999} and transmon~\cite{Koch2007} qubits. Today, it is ubiquitous in modelling superconducting circuits~\cite{tinkham, Devoret_review, Devoret_lesHouches}.
The Hamiltonian can be expressed in the mixed, phase, or charge basis:
\begin{equation}
\begin{split}
    H_T &= 4\Ec \hat{\mm}^2 - \EJ \cos \hat{\phi} \\
        &= -4\Ec \partial_\phi^2 - \EJ \cos \phi \\
        &= 4\Ec \sum_\mm \mm^2 \ket{\mm}\bra{\mm} - \frac{\EJ}{2} \sum_\mm \ket{\mm}\bra{\mm+1} + \mathrm{H. c.} 
\end{split}
\label{eq:transmon_H}
\end{equation}
The Hamiltonian is analytically solvable using Mathieu functions~\cite{Koch2007, Cottet_thesis}. This description applies to transmons of different types, including tunables ones (gatemons) \cite{Larsen2015,deLange2015,zheng2023coherent}.

The second line of Eq.~\eqref{eq:transmon_H} describes a particle in the $\phi$-space with an effective mass $m \propto 1/E_c$, trapped in a potential $V(\phi) = - E_J \cos \phi$ \cite{Clarke1988}. 
For $\Ec \ll \EJ$, corresponding to a very massive particle, the ground state is localized at the bottom of the potential well at $\phi = 0$ (for $E_J>0$, i.e., in a 0-junction), while for $\Ec \gg \EJ$, corresponding to a very low mass, the particle uniformly occupies the entire $\phi$-range. It is thus localized in the dual $m$-space: the charge difference across the junction is well defined and there are large fluctuations of the superconducting phase. In general, the wavefunction has a finite width in either space, controlled by the ratio $E_c/E_J$.

The wavefunctions of the excited states are similarly localized, but with an increasing number of peaks and nodes. This is a general property of a particle trapped in a potential well.

\subsection{Embedded QD}
 
The presence of a QD in the junction introduces pair-breaking processes and lowers the energy of quasiparticle states by binding them to the QD spin~\cite{yu1965,Shiba1969,rusinov1969,Hewson_book}.
Therefore, the approximate treatment with a model formulated in terms of Cooper pairs alone is no longer adequate and the full electron dynamics has to be considered.
In order to accurately incorporate the QD physics into the transmon equation, the basis has to be extended with the QD degrees of freedom and (this is the key point) one must allow for the presence of quasiparticles in the SCs.

The QD can alter the sinusoidal Josephson potential in complex ways, and in turn influence the properties of the transmon excitations.
The most direct example is the $\pi$-junction~\cite{Rozhkov1999, Clerk2000, Vecino2003, Choi2004, Oguri2004, vanDam2006}: if the QD is occupied by a single electron, transferring a Cooper pair across the junction requires permuting it over the QD electron. This process produces a fermionic minus sign, which effectively flips the sinusoidal potential from $-\cos\phi$ to $+\cos\phi$, so that its minimum is at $\phi = \pi$.  

For $\Ec \ll \EJ$, a good approximate approach consists of dividing the problem into two steps. 
First, one solves the impurity problem of a QD coupled to SC leads at fixed $\phi$ values to obtain the effective Josephson potentials $V(\phi)$, one for each eigensolution of the QD problem, and then uses these potentials as an input for the transmon equation where it replaces the $\cos \phi$ term~\cite{Josephson_potentials, Bargerbos2022_singletDoublet, Pita-Vidal:2022,sahu2023ground}.
However, this adiabatic approximation does not capture the possible dynamic coupling between the QD degrees of freedom contained in $V(\phi)$ and the transmon excitations. In particular, the approximation is expected to break down when two eigenstates approach each other or even cross (similar
to the breakdown of the Born-Oppenheimer approximation due to vibronic coupling in molecules).

\subsection{Reference Josephson junction}

For various applications, it is important to have the ability to impose the phase difference in the ground state~\cite{Bulaevskii1978}. 
The simplest way to achieve this is to embed the QD JJ into a larger superconducting circuit, with a second standard Josephson junction with larger Josephson energy connecting the two SCs~\cite{Vion2002, Vion2003, Koch2007, Hays2021, Pita-Vidal:2022}. Here we refer to it as the \textit{reference JJ}.
The phase difference is then controlled by piercing the resulting loop with a tunable magnetic flux. We will neglect the loop inductance and the charging energy of the reference junction, but they could be included in the model if so required.

The system containing two Josephson junctions has an effective potential \cite{DellaRocca2007,Ginzburg2018}
\begin{equation}
    V(\phi) = V_\QD(\phi) - E_J^\mathrm{ref} \cos( \phi - \phiext ),
    \label{eq:Vphi}
\end{equation}
with $E_J^\mathrm{ref}$ the Josephson energy of the reference JJ and $\phiext=2\pi \Phi/\Phi_0$ the enforced external phase due to the magnetic flux $\Phi$, with $\Phi_0=h/(2e)$ the magnetic flux quantum.
The effective potential of the QD-junction, $V_\QD$, depends on the QD parameters and is state-dependent.
Enforcing the phase is possible if the Josephson energy of the reference junction is much larger than that of the QD-junction.
Then, the minimum of $V(\phi)$, $\phi_\mathrm{min}$, is close to $\phiext$, and thus the ground state $\phi$ tends to approximately follow $\phiext$.  

\subsection{Spin-orbit coupling}

Spin-orbit terms couple spin to the supercurrent and break spin degeneracy even in the absence of the external magnetic field (except at $\phi=0$ and $\phi=\pi$, where anti-unitary symmetry leads to Kramers degeneracy). In ASQs, the matrix-valued potential energy for the doublet is given by
\begin{equation}
V(\phi)=E_0 \cos\phi - E_\mathrm{SO} \boldsymbol{\sigma} \cdot \mathbf{n} \sin \phi + g \mu_B \frac{1}{2} \boldsymbol{\sigma} \cdot \mathbf{B},
\end{equation}
where $\boldsymbol{\sigma}$ is the spin operator, $\mathbf{n}$ is a unit vector along the spin-polarization direction of SOC, $E_\mathrm{SO}$ and $E_0$ are spin-dependent and spin-independent Cooper pair tunneling rates  \cite{Bargerbos:22b}, $g$ is the g-factor, $\mu_B$ the Bohr magneton, and $B$ the magnetic field. The $g$-factor is itself $\phi$-dependent due to the impurity Knight shift \cite{knight}.

\section{Model formulation}
\label{sec:model}

\begin{figure}
    \centering
    \includegraphics[width=0.8\columnwidth]{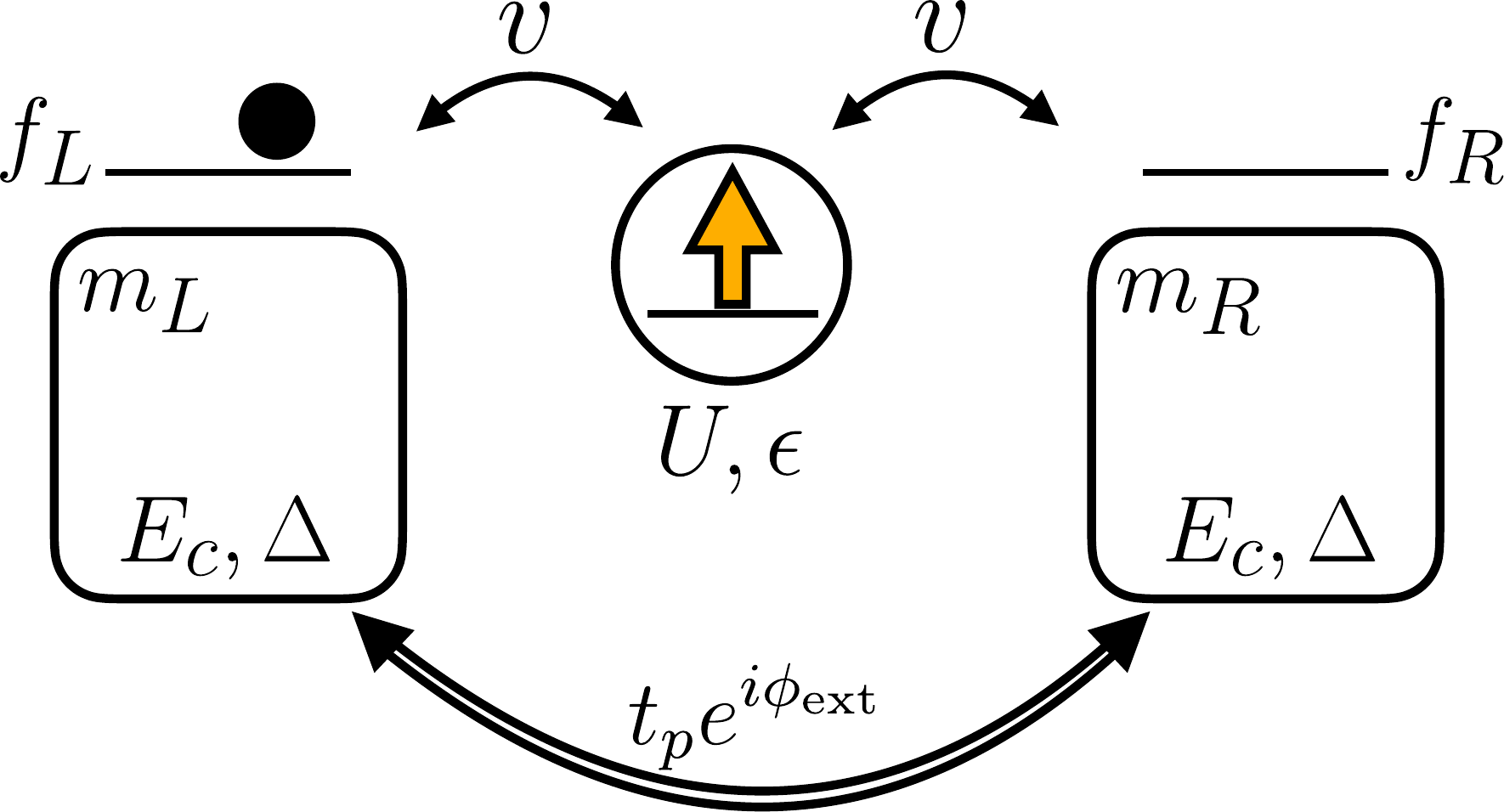}
    \caption{
    Model sketch. 
    The QD is a single energy level embedded between two superconductors in the flat-band limit of the Richardson model. 
    The QD is coupled to active superconducting orbitals $f_L$ and $f_R$ via single-electron hopping $v$.
    Far from the QD the two superconductors form another auxiliary junction, modelled by the pair-hopping $t_p e^{i \phiext}$ terms and represented by the doublet black arrow. This SQUID-like geometry enables control over the phase difference $\phi$.
    The black dot represents a possible quasiparticle occupying the $f$-orbital.
    }
    \label{fig:model_sketch}
\end{figure}

In the following we first present a model of a QD embedded between two superconducting islands, as sketched in Fig.~\ref{fig:model_sketch}.
We introduce the flat-band approximation which is necessary to reduce the Hilbert space and present the procedure to generate the reduced basis.  

\subsection{Model}

Our approach is based on describing the superconducting contacts by the Richardson model~\cite{Richardson1964,Dukelsky2000, pavesic2021}, a charge conserving pairing Hamiltonian. 
It is equivalent to the BCS mean-field theory in the thermodynamic limit \cite{Roman2002,Dukelsky2004} and encapsulates the finite-size effects observed in nanoscopic metallic grains \cite{vonDelft2001}.
Importantly, charge conservation enables a trivial implementation of the SC charging energy terms and the modeling of coupled QDs \cite{pavesic2021}.

The system consists of two SCs $(\beta = L, R)$, coupled through an embedded QD.
The full model Hamiltonian is the sum of
\begin{equation}
    \begin{split}
        H_\QD &= \epsilon n_\QD + U n_{\QD \uparrow} n_{\QD \downarrow} = \frac{U}{2} (n_\QD - \nu )^2 + \mathrm{const.}, \\
        H_\SC^{(\beta)} &= \sum_i \epsilon_i n_{\beta i} + g \frac{1}{N} \sum_{ij} c^\dagger_{\beta i \uparrow} c^\dagger_{\beta i \downarrow} c_{\beta j \downarrow} c_{\beta j \uparrow} \\
        &\quad\quad\quad\quad\quad\quad\quad\quad + \Ec^{(\beta)} \left( n_\beta - n_0^{(\beta)} \right)^2, \\
        H_\mathrm{hyb}^{(\beta)} &= \frac{v_\beta}{\sqrt{N}} \sum_{i, \sigma} \left( d^\dagger_\sigma c_{\beta i \sigma} + c_{\beta i \sigma}^\dagger d_\sigma \right), \\
        H_\mathrm{ref} &= t_p e^{i \phiext} \frac{1}{N} \sum_i c_{L i \uparrow}^\dagger c_{L i \downarrow}^\dagger \sum_j c_{R j \downarrow} c_{R j \uparrow}  + \mathrm{H. c.}
    \end{split}
    \label{eq:fullH}
\end{equation}
Here $n_{\QD \sigma} = d^\dagger_\sigma d_\sigma$ is the QD number operator and $n_\QD = n_{\QD \uparrow} + n_{\QD \downarrow}$. $\epsilon$ is the QD level and $U$ the on-site Coulomb repulsion. In the alternative formulation of $H_\QD$, $U/2$ plays the role of the QD charging energy and $\nu = \frac{1}{2} - \frac{\epsilon}{U}$ is its optimal occupation in the units of particle number (gate charge).

$\epsilon_i$ is the set of $N$ equally spaced energy levels that represent the SC, with $c_{\beta i \sigma}$ the corresponding annihilation operators, and $n_{\beta i}$ the number operators. $g$ is the superconducting pairing strength \cite{flatband}.
$\Ec^{(\beta)}$ is the SC charging energy with $n_\beta = \sum_{i} n_{\beta i}$ the total SC charge operator and $n_0^{(\beta)}$ the optimal occupation in units of particle number.
The QD and SCs are coupled by single particle hopping with strength $v_\beta$. 

The $H_\mathrm{ref}$ term describes the reference JJ.
Because it is far from the QD, it can be safely assumed that it does not contain quasiparticles, and can thus be described by the pair-hopping processes. $t_p$ is the pair hopping strength and $\phiext$ the phase difference externally imposed by the magnetic flux that pierces the superconducting loop.

We consider symmetric parameters in both superconductors and thus drop the subscript $\beta$ in the parameter symbols.

The Hamiltonian has a compact matrix product operator (MPO) representation and can in principle be solved with the density matrix renormalization group (DMRG)~\cite{twochannel}.
However, excitations in the regime we are interested in here, with $\Ec$  much smaller than the superconducting gap $\Delta$, present certain challenges.

The reason is hidden in the nature of the lowest excitations of Eq.~\eqref{eq:fullH}. 
Consider the uncoupled limit with $v = 0$, and $\Ec \rightarrow 0$. The lowest excitations are due to charge redistribution: if the ground state configuration (electron occupations of constituent parts) is $(n_L, n_\QD, n_R)$, there exists an otherwise equivalent configuration with a single shifted Cooper pair, $(n_L-2, n_\QD, n_R+2)$. Its excitation energy ($2d$ for moving a Cooper pair to a higher single-particle level) is only due to the finite level spacing, which is a consequence of the finite size of the system. 
In the thermodynamic limit ($N \rightarrow \infty$, $d \rightarrow 0$) such states are degenerate and at finite $v$ they form equal superpositions with a well-defined phase difference $\phi$.
In numerical calculations with a finite system, however, the finite excitation energy precludes a clear emergence of $\phi$.
(See Sec.~II D of Ref.~\onlinecite{twochannel} for a detailed description of the problems and limitations of the full model.)

The effect of the level spacing $d$ on the Josephson effect in the Richardson model was studied in the context of nanoscopic metallic grains in Ref.~\onlinecite{Gobert_2004}. Interestingly, they found that as $d$ is increased from zero, the effective $\EJ$ at first decreases as finite $d$ acts like a charging energy term, incurring a small energy penalty for each charge transferred. Additionally, increasing $d$ increases the pair-hopping matrix elements. At values of $d$ comparable to the superconducting gap $\Delta$ the second effect dominates, and $\EJ$ actually increases beyond the BCS value found at $d \rightarrow 0$.  

However, the goal of this work is to model superconductors in the regime where the level spacing is negligible and the charging energy comes from other effects, such as the capacitance of the constituent parts of the device.

\subsection{Flat-band approximation}

The desired behavior---the emergence of a well-defined $\phi$ from an equal superposition of degenerate $\ket{m}$ states---can be recovered by disregarding the kinetic energy of the SC levels by setting all $\epsilon_i$ to zero. This is the flat-band approximation.
We have shown in Ref.~\onlinecite{flatband} that employing this approximation does not importantly change the low-energy physics of a QD-SC system, so that the results remain  qualitatively correct.
Simultaneously, the model simplifies to the point where certain limits are analytically solvable.
Here, this treatment is extended to the SC-QD-SC setup.
In the following we present simplified expressions valid in the limit of a large number of SC levels for half-filled bands. 
\footnote{For detailed derivations and expressions at general filling see Ref.~\onlinecite{flatband}. }

Making all SC levels nominally equivalent allows us to define a single \textit{active orbital} in each SC:
\begin{equation}
    f_{\beta\sigma} = \frac{1}{\sqrt{N}} \sum_i c_{\beta i \sigma}.
\end{equation}

This greatly simplifies the hybridisation term, which now only involves the $d$ and $f$ orbitals:
\begin{equation}
    H_\mathrm{hyb}^{(\beta)} = v_\beta \sum_\sigma \left( d^\dagger_\sigma f_{\beta \sigma} + f^\dagger_{\beta \sigma} d_\sigma \right).
\end{equation}

Because the kinetic energy term is zero, the $H_\SC$ terms simplify as well. After subtracting the condensation energy of $m_\beta$ pairs ($\frac{g}{2} m_\beta$), they read:
\begin{equation}
\begin{split}
    H_\SC^{(\beta)} =& \frac{g}{2} \sum_\sigma f^\dagger_{\beta \sigma} f_{\beta \sigma} + \\
    &\quad E_c^{(\beta)} \left[  \sum_\sigma f^\dagger_{\beta \sigma} f_{\beta \sigma} + 2 m_\beta  - n_0 \right]^2.
\end{split}
\end{equation}
The pairing term simply counts the number of quasiparticles occupying the $f$-orbitals, each contributing $\frac{g}{2}$ to total energy. The SC gap $\Delta$ is thus proportional to pairing strength, which is a well-known feature of the flat-band systems~\cite{ belyaev1961,khodel1990,volovik2018}. 
Note that the approximation retains the information about the number of Cooper pairs in each SC, $m_\beta$. 
A state of the SC island in the flat-band approximation is thus fully determined by the state of the $f$-orbital (0, $\uparrow$, $\downarrow$, $2 = \downarrow\uparrow$) and $m_\beta$, the number of Cooper pairs in the condensate.

As all SC levels are equivalent, the condensate is spread equally across all of them.
The ground state of $H_{\SC}^\beta$ containing $m_\beta$ pairs is:
\begin{equation}
    \ket{m_\beta, 0} = \mathcal{N} \left( \sum_{i=1}^N c^\dagger_{\beta i \downarrow} c^\dagger_{\beta i \uparrow} \right)^{m_\beta}  \ket{0},
\end{equation}
with normalization $\mathcal{N}$~\cite{flatband} and electronic vacuum $\ket{0}$. The second label in the ket of the left hand side denotes the state of the $f$-orbital.

Excitations containing quasiparticles are
\begin{equation}
    \ket{m_\beta, \sigma} = \frac{1}{\sqrt{N}}\sum_{b=1}^N c^\dagger_{\beta b \sigma} \mathcal{N} \left( \sum_{i\neq b}^N c^\dagger_{\beta i \downarrow} c^\dagger_{\beta i \uparrow} \right)^{m_\beta}  \ket{0}.
\end{equation}
where the quasiparticle blocks the $b$-th level from participating in pairing \cite{vonDelft2001}, while the rest contain the pair condensate.
Similarly
\begin{equation}
    \ket{m_\beta, 2} = \frac{1}{N} \sum_{bb'}^N c^\dagger_{\beta b \downarrow} c^\dagger_{\beta b' \uparrow} \mathcal{N} \left( \sum_{i \neq b, b'}^N c^\dagger_{\beta i \downarrow} c^\dagger_{\beta i \uparrow} \right)^{m_\beta}  \ket{0},
\end{equation}
with $b$ and $b'$ blocked. 
The superconducting nature of the quasiparticles occupying the $f$-orbitals is contained in their algebra, given by~\cite{flatband}:
\begin{equation}
    \begin{split}
        f^\dagger_{\beta\sigma} \ket{m_\beta, 0} &= \frac{1}{\sqrt{2}} \ket{m_\beta, \sigma}, \\
        f^\dagger_{\beta\downarrow} f^\dagger_{\beta\uparrow} \ket{m_\beta, 0} &= \frac{1}{2} \left( \ket{m_\beta, 2} + \ket{m_\beta+1, 0} \right).
    \end{split}    
    \label{eq:f_algebra}
\end{equation}
Importantly, the second equation shows the possibility of recombination of two quasiparticles in the orbital $f$ into a Cooper pair. 
This is precisely the process involved in the Cooper pair transfers in a JJ.

\subsection{Full active orbital set}
\label{sec:basis}

The full basis is generated as the tensor product of eigenstates of each  subsystem for zero coupling.
The basis states are determined by the three \textit{active orbitals} $d, f_L, f_R$ and quantum numbers $\ml$ and $\mr$.
Additionally, after fixing the total charge in the system to $n$, the $(\ml, \mr)$ pair can be replaced by the difference, $\dm = \ml - \mr$.
We thus denote the basis states by
\begin{equation}
    d ~ \ket{\dm, f_L, f_R}
\end{equation}
with an operator $d = \id_d, d^\dagger_\uparrow, d^\dagger_\downarrow, d^\dagger_\downarrow d^\dagger_\uparrow$, and labels $f_{L/R} = 0, \uparrow, \downarrow, 2$. These contain all states where the quasiparticles are coupled to the QD~\cite{flatband}. This step beyond basis sets that only involve Cooper pairs \cite{matute2023} is necessary in the presence of magnetic-impurity-induced pair-breaking processes.

Furthermore, our model conserves the total spin $S$. 
We consider the singlet ($n = \mathrm{even}$, $S=0$) and the doublet ($n = \mathrm{odd}$, $S=1/2$, $S_z = +1/2$) sectors.
Because Cooper pairs are singlets, $S$ is solely determined by the state of the three active orbitals.
For each $\dm$, there are 14 possible states in each sector. 
(See App.~\ref{appA} for their definitions.)

We generate the matrix representation of the Hamiltonian by computing symbolic expressions for all matrix elements for general $[\ml, \mr]$ using symbolic algebra software \cite{Mathematica} and then selecting the allowed $\dm$ configurations for a given total charge $n$.
The basis size grows linearly with $n$, which enables efficient exact diagonalization for $n$ in the thousands. 
The computer code for performing numerical calculations is available in a public repository \cite{flat2023}.

\subsection{Fourier transform to the phase-space }

The phase difference $\phi$ and the difference in Cooper pair number $\dm$ are conjugate quantities~\cite{tinkham}, i.e., $[\phi,\dm]=i$, and thus related via Fourier transform. 
When performed on the $\dm$-states, we obtain a basis in the $\phi$-space:
\begin{equation}
    \ket{\phi, f_L, f_R} = \frac{1}{\sqrt{\mtot+1}} \sum_{\dm} e^{i \phi \left( \dm + \mtot \right)/2} \ket{\dm, f_L, f_R},
    \label{eq:FT}
\end{equation}
with $\mtot$ the total number of Cooper pairs in the system. $\dm$ runs across all possible configurations, $\dm = -\mtot, -\mtot + 2, \dots, \mtot-2, \mtot$; there are $\mtot+1$ such configurations.
The multiplicative factor of $1/2$ in the exponent is necessary because of the steps of 2. Furthermore, we shift $\dm$ by $\mtot$ for convenience. The phase $\phi$ is defined in the interval $[0, 2\pi)$:
\begin{equation}
    \phi = 2\pi \frac{l}{\mtot+1},
\end{equation}
with $l=0,\ldots,\mtot$.
In the thermodynamic limit, $N\to\infty$, $\phi$ becomes a periodic continuous variable \cite{Devoret2021}.

\begin{figure}
    \centering
    \includegraphics[width=\columnwidth]{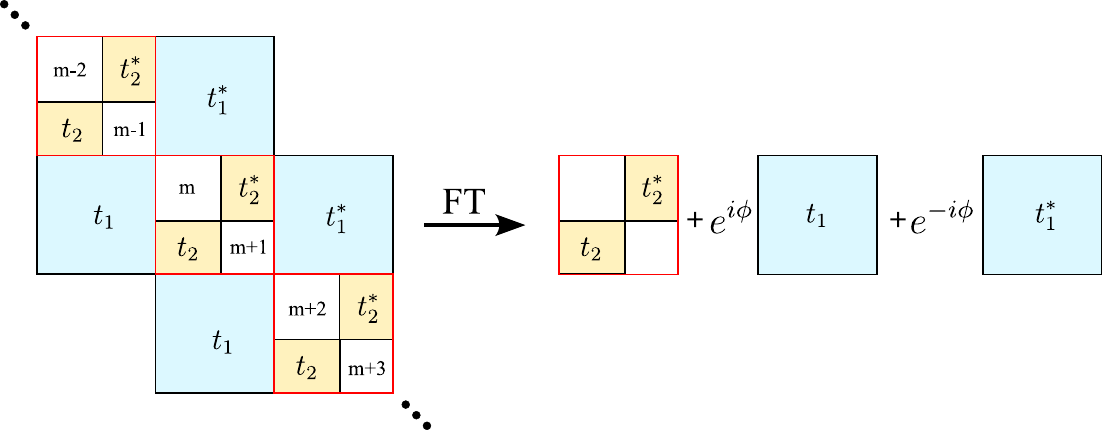}
    \caption{
    Block diagonalizing the Hamiltonian at $\Ec=0$ with the Fourier transform.
    In the analogy with the tight-binding chain, the $t_2$ block denotes coupling within each chain site, while $t_1$ is hopping between the sites. 
    }
    \label{fig:FT}
\end{figure}

For $\Ec = 0$ and neglecting finite-size effects, $\phi$ is a good quantum number. 
Therefore, in this case the Hamiltonian---a block tridiagonal matrix in the $\dm$-basis---has a block-diagonal shape when expressed in the $\phi$-basis, with the matrix elements inside the block describing the internal degrees of freedom associated with the full active orbital set (unpaired particles), see Fig.~\ref{fig:FT}.
These are $14 \times 14$ matrices, with the phase variable $\phi$ appearing in the blue blocks of hopping matrix elements, similar to what happens in the Bloch state basis in the tight-binding description of electrons on a lattice. We remind the reader that the matrices are different in the singlet and doublet sectors.
By diagonalizing them we obtain the eigenstates in the $\phi$-basis.

\subsection{Charge vs. phase basis: Similarities and differences compared to the zero-bandwidth BCS}

The model resembles the zero-bandwidth BCS approximation (ZBW BCS), which is a popular tool for qualitatively describing the physics of hybrid superconducting QD devices \cite{affleck2000, Vecino2003, bergeret2007,GroveRasmussen2018,Zonda2022, Hermansen2022, Schmid2022}.
Indeed, in the simplest case of a QD coupled to a single SC, the two models are actually mathematically equivalent \cite{flatband}.
However, we argue that the flat-band Richardson formulation presented here is better suited for modelling extended quantum devices with possibly complex geometries, and where charging energies of the constituent parts play an important role. 

An important aspect is that the degrees of freedom associated with the Cooper pair condensate do not appear explicitly in the ZBW BCS, while in the flat-band Richardson model we keep track of them with the labels $m_\beta$. While this is not important for a single SC, the condensate plays a key role when two (or more) SCs are coupled; as discussed just above, the transfers of Cooper pairs (changes of $m=m_L-m_R$ by $\pm 2$) generate the Josephson effect and lead to the emergence of the phase difference $\phi$.
This is a statement equivalent to the notion that the phase of a single isolated SC is arbitrary (gauge freedom), but the \textit{phase difference} between two coupled SCs is physically measurable and gauge invariant.

These details are not immediately obvious in the BCS model which is formulated in the phase basis, with $\phi$ taking a fixed value due to the gauge symmetry breaking.
This becomes an issue, however, when considering non-zero charging energy. 
In that case $\phi$ is not a good quantum number, and thus cannot be represented by a constant value, but needs to be promoted to an operator quantity in order to allow for the phase fluctuations in the model.
The problem is naturally resolved in the charge conserving formalism of the Richardson model, where $\phi$ is an emergent global degree of freedom, arising as a property of a superposition of many $\ket{m}$ states. This means that the model is able to accurately capture its quantum fluctuations at finite $\Ec$.

Alternatively, it is possible to describe the problem in the $\phi$-basis, while also retaining information about the pair condensate. Indeed, this is achieved by the Fourier transform in Eq.~\eqref{eq:FT}. 
In that case the charging energy terms appear as off-diagonal matrix elements which couple different $\ket{\phi}$-states.

\subsection{Magnetic field and spin-orbit coupling}

The method can be easily extended to include additional effects. For example, the magnetic field on the QD along the $z$ axis is included by substituting $\epsilon \rightarrow \epsilon_\sigma = \epsilon + \sigma E_Z/2$, where $E_Z$ is the Zeeman splitting of the QD level, $E_Z=g\mu_B B$, and $\sigma = \pm 1$, depending on the electron spin.
The $x$ and $y$ components of the magnetic field appear as off-diagonal elements coupling the $S_z = 1/2$ and $S_z = -1/2$ basis states. 

The spin-orbit coupling can be included in similar way as in Ref.~\onlinecite{Bargerbos:22b}, by adding two terms: a spin-flip SC-QD hopping $v_{\uparrow \downarrow}$ and a SC-SC single-electron hopping $t_\mathrm{sc}$ (see also the Supplementary material of Ref.~\onlinecite{Bargerbos:22b}):
\begin{equation}
\begin{split}
    \sum_{\sigma} \left( i v_{\uparrow \downarrow} d^\dagger_\sigma f_{L \bar{\sigma}} +
    i v_{\uparrow \downarrow} f^\dag_{R\sigma} d_\sigma +  \mathrm{H. c.} \right) \\
    +  \sum_\sigma \left( t_\mathrm{sc}  f_{L \sigma}^\dagger f_{R \sigma} + \mathrm{H. c.} \right).
    \end{split}
    \label{eq:soc}
\end{equation}
Here $\bar{\sigma}$ indicates the reversed spin index. With this definition, the spin-polarization direction of SOC is along the $x$-axis. The field along the $x$ axis is said to be ``parallel'', while fields along the $y$ and $z$ axes are said to be ``perpendicular''.

\section{Benchmark results}
\label{sec:results}

In this section, we demonstrate that the presented model reproduces the expected results in appropriate limits. We also point out the existence of parameter regimes where the generalized formalism reveals less familiar behavior.

\subsection{Role of quasiparticles}

In our formalism, four electron hopping events are necessary for a Cooper pair to pass through the QD, and thus we expect $\EJ$, the half-width of the cosine-like excitation band, to increase proportionally to $v^4$.
However, $\phi$-dependent processes of second order in $v$, when present, will dominate the dynamics of the junction in the perturbative regime.
Here we show that this happens in the presence of a superconducting quasiparticle and identify the parameter ranges where such contributions become large due to the presence of the QD.
In such cases the treatment with an effective pair-hopping Hamiltonian is not adequate, and solving the QD-JJ Hamiltonian on the level of single-electron processes is necessary. 

\subsubsection{Subgap spectrum of the SC-QD-SC junction}

\begin{figure}
    \centering
    \includegraphics[width=\columnwidth]{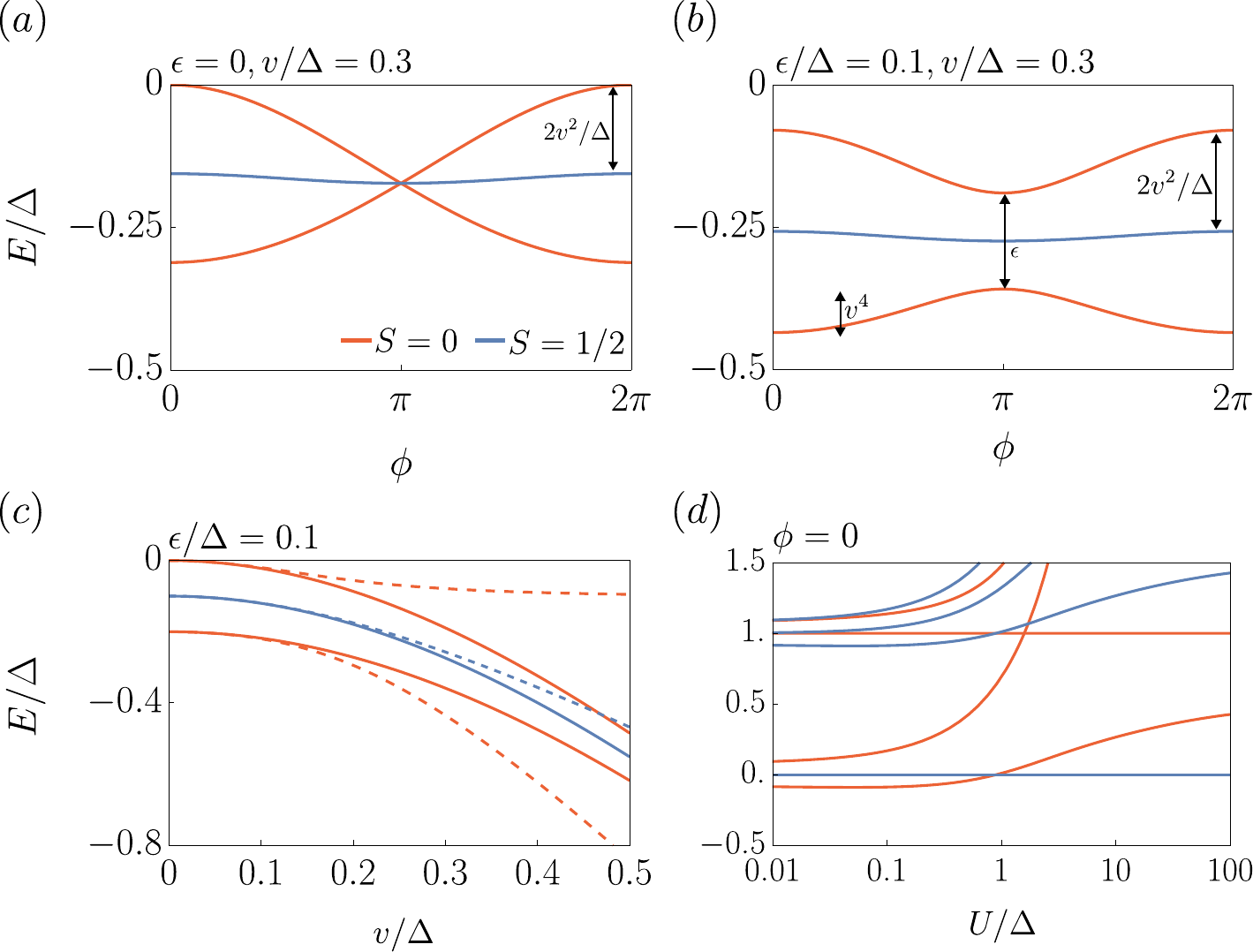}
    \caption{
    Spectra of the SC-QD-SC junction for $\Ec=0$, $t_p=0$. 
    \textbf{(a, b)} $\phi$ dependence of the lowest  singlet (red) and the lowest doublet (blue) states for (a) the resonant case of $\epsilon = 0$ and (b) for the generic case of $\epsilon = 0.1\Delta$. Here $U=0$.
    \textbf{(c)} $v$ dependence of the two lowest singlets and the lowest doublet for $\phi = 0$ (dashed) and $\phi = \pi$ (solid), for $U=0$.
    \textbf{(d)} $U$ dependence of the spectra at $\phi=0$ close to the particle-hole symmetric point, $\epsilon = -0.8 U/2$. The energy zero coincides with the energy of the lowest doublet. Here we use $v/\Delta = 0.2 (1+\sqrt{U/\Delta})$, because the main goal is to study the effect of the $U/\Delta$ ratio, while keeping the spin exchange interaction $v^2/U$ constant in the large-$U$ regime.
    }
    \label{fig:spectra}
\end{figure}

We begin by reproducing the basic properties of a JJ with an embedded QD, starting with $t_p=0$, i.e., without the reference junction, and $E_c=0$, i.e., without any charging energy terms. In this simplest case the phase difference $\phi$ is a conserved scalar quantity (good quantum number) that labels the eigenstates of the system.

We start by showing the $\phi$ dependence of the two lowest singlets and the lowest doublet for a non-interacting ($U=0$) QD in Fig.~\ref{fig:spectra}(a,b).
The SC leads induce superconducting pairing on the QD level, which results in a $\phi$-dependent splitting of the states according to the occupation of the QD.
This is known as the proximity effect.
In the doublet state the QD is occupied by a single electron, while in the singlet states it contains a Cooper pair or two Bogoliubov quasiparticles (broken Cooper pair), respectively.
In the electron occupation basis, these are represented as orthogonal equal superpositions of $\ket{0}$ and $\ket{2}$ at the QD. (For $\phi = 0$, we have $\ket{0}+\ket{2}$ for a Cooper pair, and $\ket{0}-\ket{2}$ for a broken pair.)

The resonant situation ($\epsilon=0$),  Fig.~\ref{fig:spectra}(a), shows some particularities. The singlets are split through a second order $v^2$ process where a virtual quasiparticle hops from the QD to a SC orbital and back.
This SC orbital is equally distributed across the two SCs (assuming symmetric $v = v_L = v_R$), with the phase between the contributions of the two leads corresponding to the $\phi$-dependent symmetry of the system~\cite{qdjj}.
This results in a $\phi$-dependent correction of energy $(2 v^2 / \Delta) \cos(\phi/2)$~\cite{Vecino2003}. The resulting $4\pi$ periodicity of the energy dispersion in the singlet state and the quadratic dependence on $v$ are the main signatures of resonant behavior.

In Fig.~\ref{fig:spectra}(b) we show the generic (non-resonant, particle-hole asymmetric) case with non-zero $\epsilon$. 
This regime corresponds to a standard JJ with subgap Andreev bound states (ABS)~\cite{Andreev_reflection, tinkham}.
In comparison to the resonant case shown in Fig.~\ref{fig:spectra}(a), here the particle-hole asymmetry splits the singlets at $\phi=\pi$. 
The dispersion is then $2\pi$-periodic.
The ground state has a larger contribution of the empty QD state $\ket{0}$ (for $\epsilon>0$), while the excited singlet state has a mostly doubly occupied QD, state $\ket{2}$.
The previously discussed second order process is energetically suppressed, for it requires the formation of Cooper pairs -- equal contributions of $\ket{0}$ and $\ket{2}$.
The $\phi$-dependence of singlet states is thus dominated by the next higher order ($\propto v^4 \cos \phi$) contribution, coming from coherent transport of two electrons---a Cooper pair---across the junction.
These are the processes that generate the Josephson supercurrent, and we find the Josephson energy $E_J \propto v^4$.

Figure~\ref{fig:spectra}(c) shows the $v$-dependence of the ABS, illustrating the increase of $E_J$ that is quantified by the spread between the dashed ($\phi=0$) and solid ($\phi=\pi$) lines.
Note the significant difference between the doublet (blue) and singlet (red) states.

In Fig.~\ref{fig:spectra}(d) we show the evolution of the spectra with increasing $U$ while staying close to the particle-hole symmetric point by setting $\epsilon = - 0.8 U/2$, so that the average QD occupancy is approximately one. In the lowest doublet state the QD contains a single spin already at $U=0$. This state remains largely unperturbed for increasing $U$. In contrast, the singlet states strongly depend on the interaction strength.
The lowest singlet state gradually transforms from the ABS at $U \rightarrow 0$ to a Yu-Shiba-Rusinov (YSR) state for $U\gg\Delta$, where the electron in the singly-occupied QD forms a bound state with a Bogoliubov quasiparticle in the superconductor through the exchange interaction.
The first excited singlet retains large contributions of the $\ket{0}$ and $\ket{2}$ QD states and is thus quickly pushed to higher energies with increasing charge repulsion $U$~\cite{twochannel}.

\subsubsection{Effective Josephson energy}

In analogy with a standard JJ, we define the effective Josephson energy as the half-width of the energy band obtained by varying $\phi$,
\begin{equation}
\label{eq:EJeff}
    \EJeff = \frac{1}{2} \left( \max_\phi E(\phi)- \min_\phi E(\phi) \right),
\end{equation}
where $E(\phi)$ is the energy of the lowest state in the singlet or doublet subspace (the result is  state-dependent). The extreme values are taken at $\phi=0$ and $\phi=\pi$ when $t_p=0$.

\begin{figure}[htbp]
    \centering
    \includegraphics[width=\columnwidth]{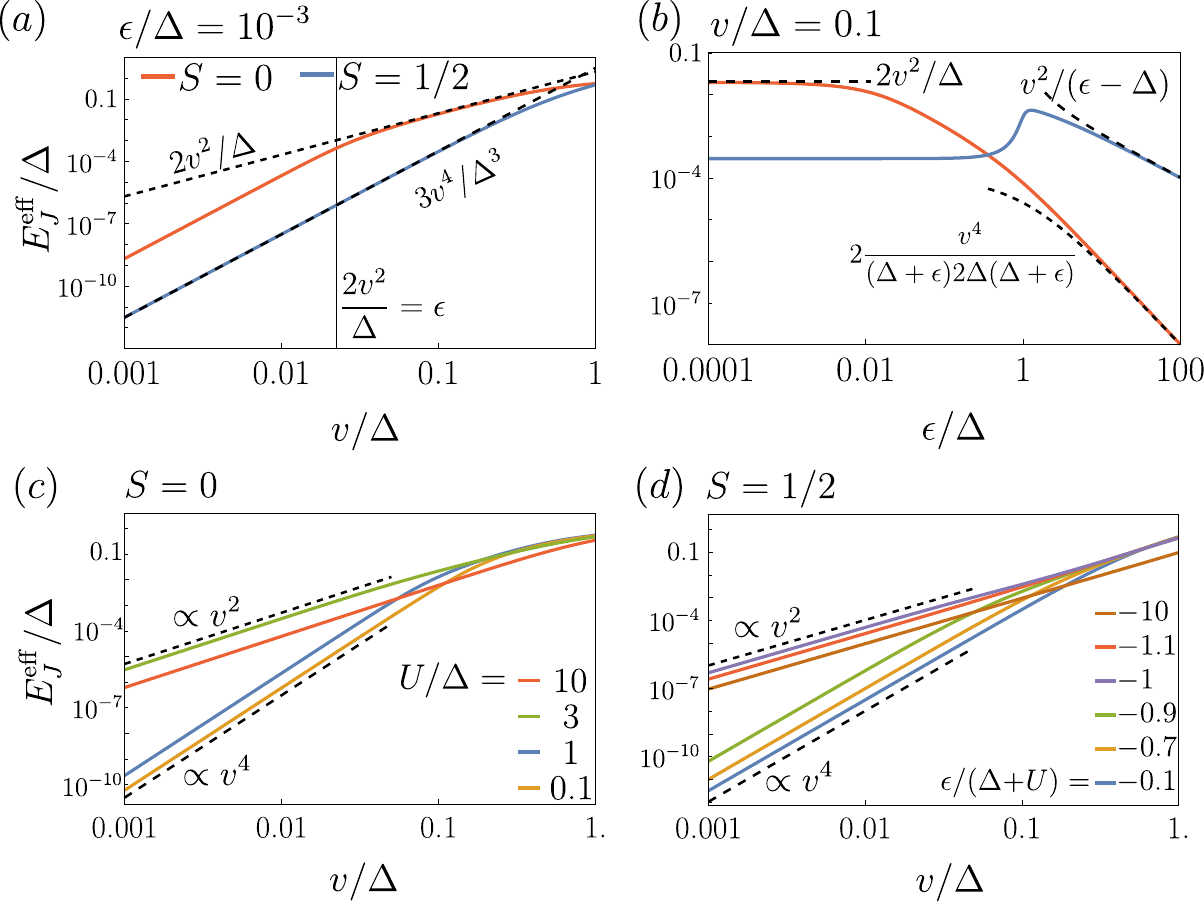}
    \caption{
    Effective Josephson energy $\EJeff$, defined as the difference in the energy of the lowest state at $\phi = 0$ and $\phi = \pi$. 
    \textbf{(a)} and \textbf{(b)} $U=0$ results. Full lines: singlet (red) and doublet (blue); black dashed lines: expected perturbative corrections.
    \textbf{(c)} $\EJeff$ for the singlet ground state for different $U$ with $\epsilon = -0.8(U/2)$, sightly shifted away from the particle-hole symmetric point.
    \textbf{(d)} $\EJeff$ for the doublet ground state with increasingly negative $\epsilon$, at $U/\Delta=0.2$. $-\epsilon = \Delta + U$ is the point where a quasiparticle becomes present in the superconductor.
    }
    \label{fig:Ejeff}
\end{figure}

Fig.~\ref{fig:Ejeff}(a) shows a log-log plot of the $v$ dependence of $\EJeff$ at small $\epsilon=10^{-3}$, with $U=0$. 
For the doublet ground state (blue) we find $\EJeff \propto v^4$ for all $v$, as the spin in the QD is not coupled to the superconducting leads and the transfer of Cooper pairs across the junction remains the leading contribution for $v < \Delta$.
(The presence of the QD spin does however change the prefactor of the Josephson current, a phenomenon known as the $\pi$-junction.) 

However, in the singlet ground state (red) we find two regimes, with $\EJeff \propto v^2$ for large $v$ and $\EJeff \propto v^4$ for small $v$.
This is a consequence of the competition between $\epsilon$ splitting the $\ket{0}$ and $\ket{2}$ QD states and the proximity effect hybridizing them into Cooper pairs.
The crossover occurs at $\epsilon = 2v^2 / \Delta$ (vertical black line). The right hand side of the equality corresponds to the hybridisation coming from the proximity effect.
The regime where $\epsilon < \frac{2v^2}{\Delta}$ qualitatively corresponds to Fig.~\ref{fig:spectra}(a), while an example of the $\epsilon > \frac{2v^2}{\Delta}$ case is shown in Fig.~\ref{fig:spectra}(b)

Fig.~\ref{fig:Ejeff}(b) shows the $\epsilon$ dependence of $\EJeff$.
For the singlet state (red) we see the evolution from the $\epsilon$-independent $v^2$ regime to the $v^4$ dependence as discussed above. 
Interestingly, by increasing $\epsilon$ beyond $\Delta$, we observe $\propto v^2$ behavior in the doublet state (blue) as well.
This is a regime where it becomes favorable for the unpaired electron from the QD to enter the SC as a quasiparticle.
Similarly to the second order process that causes the proximity effect in the singlet sector, here the quasiparticle occupies an orbital symmetrically distributed across the two SCs, and the second order $\phi$-dependent correction comes from processes where the quasiparticle traverses the junction (or conversely, from the $\phi$-dependent shape of the orbital).

We have thus established a generic property of the system that can only be captured by accounting for single electron hopping processes: the presence of a SC quasiparticle always leads to a dominant second-order contribution to $\EJeff$. 
This is further corroborated in Fig.~\ref{fig:Ejeff}(c,d). 
In the singlet ground state, Fig.~\ref{fig:Ejeff}(c), a quasiparticle is induced by increasing $U$ beyond $\Delta$, transforming the nature of the ground state from ABS to YSR. Correspondingly, $\EJeff(v)$ changes from $v^4$ for $U < \Delta$ to $v^2$ for $U > \Delta$.
Similarly, a quasiparticle is favored in the doublet ground state if $\epsilon > \Delta$ (for $\epsilon > 0$), or $\epsilon > \Delta + 2\epsilon + U$ ($\epsilon < 0$). 
Fig.~\ref{fig:Ejeff}(d) shows the negative $\epsilon$ case, where the crossover from $v^4$ to $v^2$ behavior occurs when $\vert \epsilon \vert > \Delta + U$.

\subsection{Reference Josephson junction: \texorpdfstring{$t_p$}{tp} dependence}
\label{sec:tp}

Next, we investigate the effect of the reference Josephson junction in the SQUID geometry as a means of enforcing the phase bias $\phi = \phiext$ across the QD JJ using external magnetic flux \cite{Ginzburg2018}. The Josephson energy of the reference junction is given by $\EJref = 2 t_p$. The effective Josephson energy of the isolated QD junction is denoted $\EJeff(0)$ and corresponds to $\EJeff$, defined in Eq.~\eqref{eq:EJeff}, at $t_p = 0$. The total potential is the sum of the QD-induced potential and the $-\EJref \cos(\phi -\phiext)$ term, as in Eq.~\eqref{eq:Vphi}. We note that this setup corresponds to the ASQ embedded in a transmon that was experimentally explored in Refs.~\onlinecite{Bargerbos2022_singletDoublet,Bargerbos:22b,Pita-Vidal:2022}. 

\begin{figure*}[htbp]
    \centering
    \includegraphics[width=0.99 \textwidth]{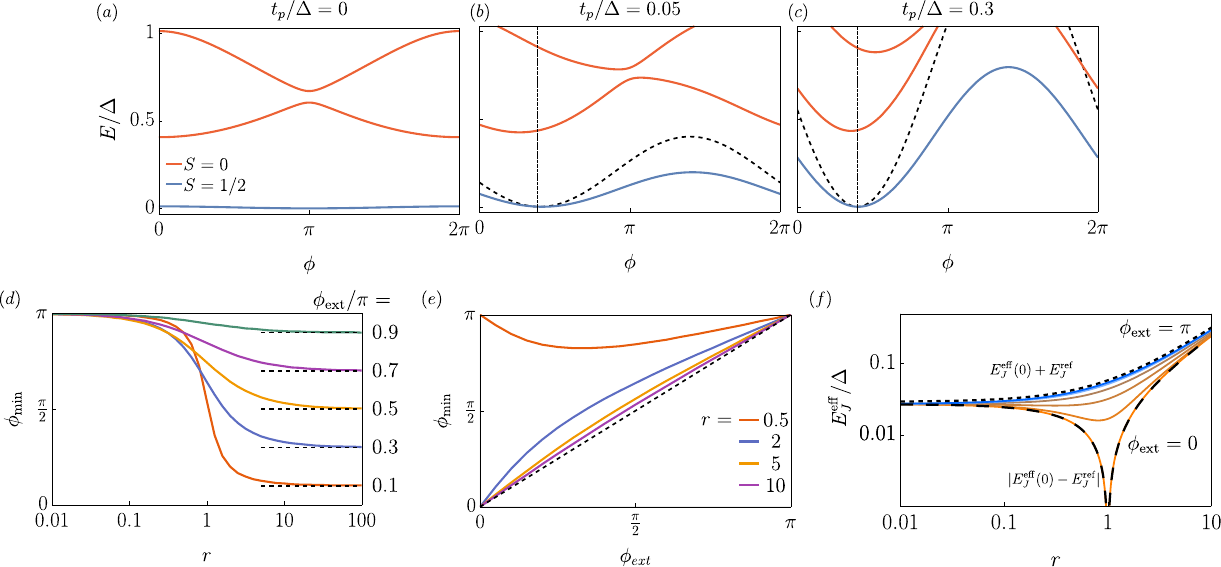}
    \caption{
    Effect of the tunneling amplitude of Cooper pairs across the reference junction, $t_p$. 
    \textbf{(a-c)} $\phi$-dependence of the spectra for three increasing values of $t_p$, at fixed $\phiext = 0.4 \pi$ (indicated with vertical lines). The black dashed curves show the Josephson potential of the auxiliary Josephson junction, $-2 t_p \cos(\phi-\phiext)$.
    \textbf{(d)} Location of the minimum of the ground state for different $\phiext$ as a function of the ratio of the Josephson energy of the auxiliary and the QD-junction $r = \EJref/\EJeff(0)$. Dashed black lines indicate the corresponding $\phiext$.
    \textbf{(e)} 
    Position of $\phi_\mathrm{min}$ vs. $\phiext$ for different values of $r$.
    \textbf{(f)}
    The black dashed line corresponds to $\phi_\mathrm{min} = \phiext$.
    $\EJeff$ vs. $r$, with $\phiext$ ranging from $0$ (orange) to $\pi$ (blue) in steps of $0.2\pi$. The top black dashed line corresponds to the sum of contributions from both junctions, $\EJeff(0) + \EJref$, while the bottom one is the absolute difference $\left\vert \EJeff(0) - \EJref \right\vert$.
    Other parameters: $U/\Delta=3$, $\epsilon = -0.9\frac{U}{2}$, $v/\Delta=0.4$, $\Ec = 0$.
    }
    \label{fig:tp_figure}
\end{figure*}

We set $U = 3\Delta$ and remain close to the particle-hole symmetric point, a parameter regime relevant for the ASQ, where the ground state is a doublet with a singly occupied QD, $\EJeff \propto v^4$, and $V(\phi)$ has a minimum at $\phi_\mathrm{min} = \pi$.
Figs.~\ref{fig:tp_figure}(a-c) show the spectra with increasing $t_p$ for $\phiext = 0.4\pi$. 
The phase difference in the ground state, $\phi_\mathrm{min}$, depends on the competition between the reference and the QD junction, which we quantify by the ratio
\begin{equation}
    r = \frac{\EJref}{\EJeff(0)}.
\end{equation}
Fig.~\ref{fig:tp_figure}(d) shows the $r$ dependence of $\phi_\mathrm{min}$ as $t_p$ is increased, and the systems transitions from a QD-induced $\pi$-junction to the one with enforced $\phi_\mathrm{min} = \phiext$. 
There is weak $\phiext$ dependence of $\phi_\mathrm{min}$ at finite $r$: the phase difference is not equally strongly enforced for all $\phiext$.

This is explicitly shown in Fig.~\ref{fig:tp_figure}(e), where we plot the location of the potential minimum $\phi_\mathrm{min}$ with changing $\phiext$ for various values of $r$.
For experimentally relevant $r \sim 5-10$ we find relatively small deviation from the optimal $\phiext=\phi_\mathrm{min}$ situation (black dashed line).

The depth of the effective total potential $V(\phi)$, defined in Eq.~\eqref{eq:Vphi}, depends on $\phiext$.
This is shown in Fig.~\ref{fig:tp_figure}(f), where we plot the $r$ dependence of the effective Josephson energy $\EJeff$ from Eq.~\eqref{eq:EJeff}.
$V(\phi)$ is a sum of two sinusoidal terms [see Eq.~\eqref{eq:Vphi}], one coming from the QD with a state-dependent phase factor ($0$ for singlet, $\pi$ for doublet) and the other from the auxiliary junction with the phase $\phiext$.
When the two phases are aligned at $\phiext = \pi$, the amplitude of $V(\phi)$ is the sum of the two amplitudes (upper black dashed line).
On the other hand, when the two junctions are out of phase at $\phiext = 0$ their contributions subtract (lower black dashed line).
At $r=1$ this leads to a completely flat $V(\phi)$ (zero $\EJeff$) despite the Josephson energies of the two junctions possibly being very large.
Any finite value of $\Ec$ close to this point would cause large fluctuations of $\phi$. 

\subsection{QD-transmon: charging energy effects}

After demonstrating that the model reproduces a number of standard properties of a QD JJ, we turn our attention to the case of finite charging energy. 
$\Ec$ induces mixing between the states with a different $\phi$, and it is no longer possible to find eigenstates in each $\phi$ subspace separately.
We thus use the full basis as defined in Sec.~\ref{sec:basis} and App.~\ref{appA}.

\begin{figure*}
    \centering
    \includegraphics[width=0.85 \textwidth]{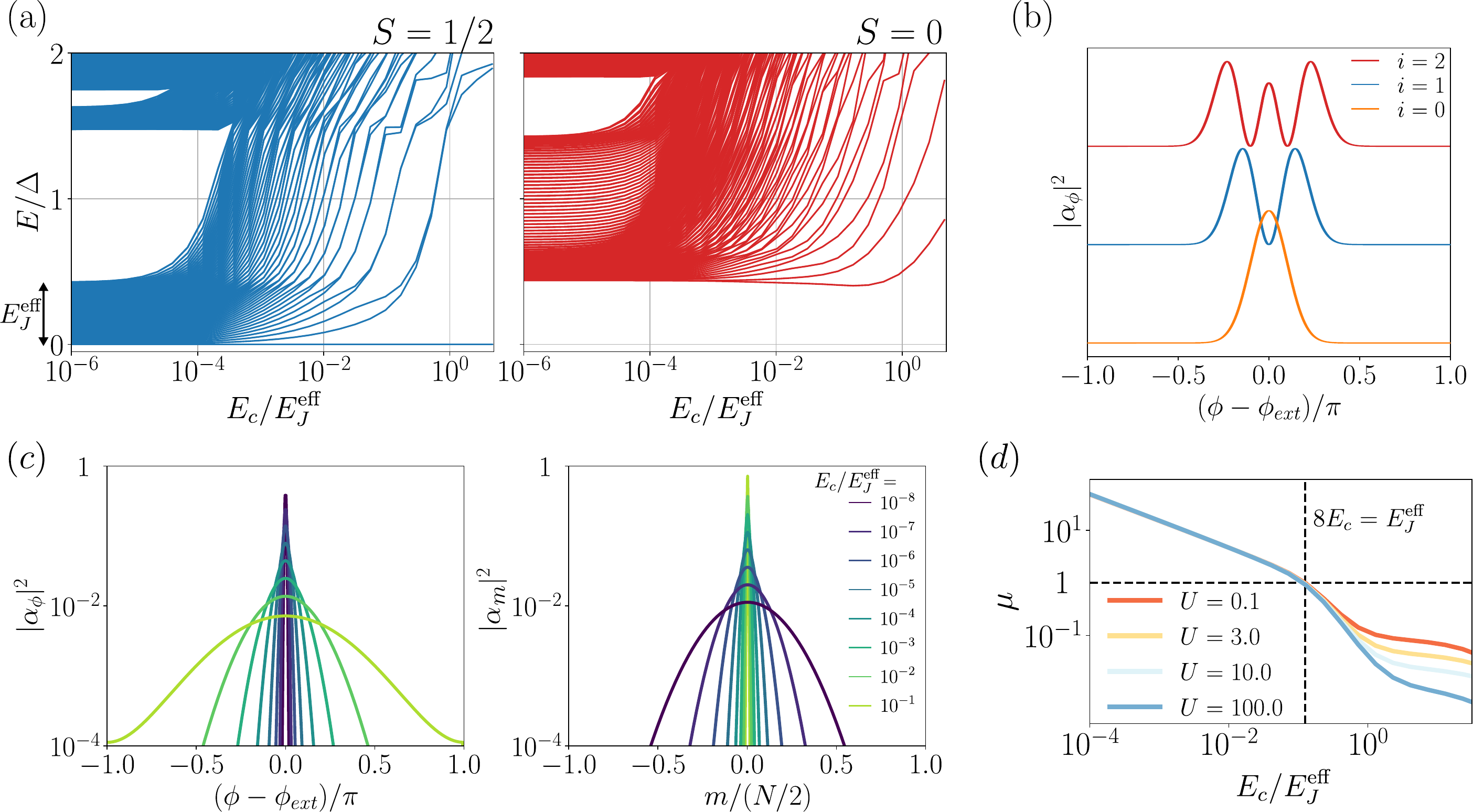}
    \caption{
    Effect of the charge repulsion $\Ec$. 
    \textbf{(a)}
    Evolution of the excitation spectrum with increasing $\Ec$ in the doublet (blue) and the singlet (red) sector.
    \textbf{(b)}
    Absolute values of the amplitudes in the $\phi$-basis for the ground state and first two excitations in the doublet sector in the intermediate regime, $\Ec / (\EJeff/2) = 10^{-3}$. 
    \textbf{(c)}
    Absolute values of amplitudes in the $\phi$-basis ($\alpha_\phi$) and $\dm$-basis ($\alpha_{\dm}$) for the doublet ground state for different values of $\Ec$. 
    \textbf{(d)}
    Variance of $\dm$, $\mu = \langle \dm^2 \rangle - \langle \dm \rangle^2$ in the doublet ground state vs. $\Ec/\EJeff$ at $\phiext = \pi$. The black vertical line corresponds to $8 \Ec = \EJeff$.
    $\EJeff(0)/\Delta = 0.027$, $r=1.1$. 
    Other parameters are $U/\Delta = 3$, $\epsilon = -0.8 \frac{U}{2}$, $v/\Delta = 0.5$, $N = 601$ and $n_0 = 300$. In (a-c) $t_p/\Delta = 0.1$, $\phiext = \pi$.
    }
    \label{fig:spectra_vs_Ec}
\end{figure*}

The effect of $E_c$ on the spectrum is shown in Fig.~\ref{fig:spectra_vs_Ec}(a). 
With increasing $\Ec$ the eigenstates lose their well-defined phase and gradually transform into states with a well-defined $\dm$.
This is reflected in the disassociation of the band of width $2\EJeff$ into discrete states with excitation energy growing as $4 \Ec \dm^2$.
In agreement with the effective transmon model (Eq.~\ref{eq:transmon_H}), the crossover occurs over a protracted range of $E_c$ values, centered at $\Ec/\EJeff \sim 10^{-3}$ .

The nature of the eigenstates can be gauged by tracing over the active orbital degrees of freedom and plotting the amplitudes in the $\dm$-basis ($\alpha_{\dm}$) and in the Fourier transformed $\phi$-basis ($\alpha_\phi$), see Fig.~\ref{fig:spectra_vs_Ec}(c).
In the $\Ec \rightarrow 0$ limit the eigenvectors tend to a $\delta$ peak in the $\phi$-space, corresponding to a wide distribution in the dual $\dm$-space. The opposite is found in the $\Ec\to\infty$ regime. 
In the crossover regime the distribution of $\alpha$ exhibits substantial width in both basis spaces. This indicates that neither $\phi$ nor $\dm$ are well defined, and there is no privileged basis for the description of eigenstates.

In Fig.~\ref{fig:spectra_vs_Ec}(b), we show $\vert \alpha_\phi \vert^2$ for the doublet states in the transmon regime, exhibiting an increasing number of peaks and nodes in higher-lying excitations. These curves can be interpreted as effective wavefunctions of the transmon degree of freedom, exhibiting the expected characteristics of a particle trapped in a potential well.

The quantum fluctuations of $\dm$ reflect the competition between the charging energy $\Ec$, favoring states with a well defined $\dm$, and the effective potential $V(\phi)$, favoring a state with a well defined $\phi$. 
In Fig.~\ref{fig:spectra_vs_Ec}(d) we plot the variance of $\dm$ in the doublet ground state,
\begin{equation}
    \mu = \langle \dm^2 \rangle - \langle \dm \rangle^2,
\end{equation}
for different values of $U$.
$\mu$ decreases with increasing $\Ec$, as a certain charge distribution becomes favored and the probabilities of other $\dm$ values are reduced.
Plotting in log-log scales uncovers two regimes. At small $\Ec$ we find roughly $\mu \propto 1/\sqrt{\Ec}$, independent of the value of $U$.
In this regime the charging energy provides a large effective mass $m^*$ in the harmonic-oscillator picture, $m^* \propto 1/E_c$, which then leads directly to $\mu \propto m^*\omega \propto \sqrt{m^*} \propto 1/\sqrt{E_c}$.

The inflection point is at $8 \Ec = \EJeff$ (dashed vertical line).
Here $8\Ec$ corresponds to the charging energy penalty of the excited state with $\dm = \pm 2$, obtained by transferring one Cooper pair across the junction; the sum of the $\Ec n^2$ terms in each SC gives $\Ec \left( 2^2 + (-2)^2 \right) = 8 \Ec$.
Notably, at this point $\mu \sim 1$, indicating a close to equal superposition of the $\ket{\dm=0}$ and $\ket{\dm=\pm2}$ states.

At larger $\Ec$, only the $\ket{\dm=0}$ state remains populated. 
Furthermore, the electrostatic effects in the QD start to play a role, as seen by the dependence of $\mu$ on $U$. 
For large $U$, the fluctuations of charge on the QD are strongly prohibited, and thus $\mu$ decreases much faster with increasing $E_c$ than for smaller values of $U$~\cite{pavesic2021}.

\subsection{Spin-orbit splitting of the doublet states}

\begin{figure}[htbp]
    \centering
   \includegraphics[width=0.65 \columnwidth]{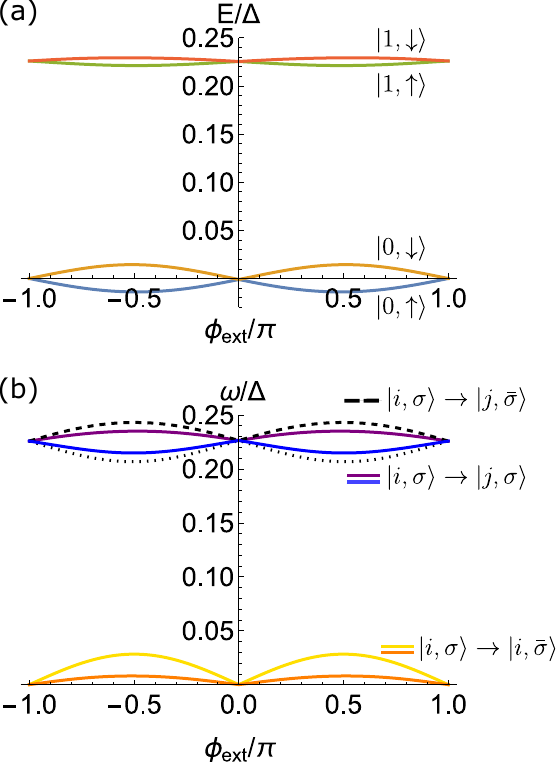}
    \caption{
    Spin-orbit splitting of the doublet states at zero magnetic field. We plot (a) the eigenenergies $E$ of the states $\ket{0,\sigma}$ and $\ket{1,\sigma}$, where $\sigma$ is the helicity index, and (b) the corresponding transition frequencies $\omega$ between these states.
    The line styles and colors for transitions resemble those from Figs.~4(a,b) in Ref.~\onlinecite{Pita-Vidal:2022}. 
    Blue and purple: pure transmon transitions. Yellow and orange: pure spin-flip transitions. Dashed and dotted: mixed transitions involving both spin and transmon degrees of freedom.
    Parameters: $U/\Delta=3$, $\epsilon = -\frac{U}{2}$, $v/\Delta=0.5$, $v_{\uparrow\downarrow}/\Delta=0.2$, $v_\mathrm{sc}/\Delta=0.2$, $t_p/\Delta=0.1$, $E_c/\Delta = 0.02$.
    }
    \label{fig:splitting}
\end{figure}

If the QD is in the doublet state, the transmon excitations (labelled by an integer $i=0,1,2,\ldots$) split into $\ket{i, \uparrow}$ and $\ket{i, \downarrow}$ states due to the Zeeman splitting of the QD level in magnetic field, or due to SOC-induced spin splitting for general values of $\phiext$. 
Figure~\ref{fig:splitting} shows the low-energy doublet states $\ket{i,\sigma}$ for the two lowest transmon levels $i=0,1$ in the presence of the SOC at zero magnetic field, as well as all six transitions between these four states. This figure can be compared with Fig.~4b in Ref.~\onlinecite{Pita-Vidal:2022}, showing the two-tone spectroscopy of the joint two-qubit (transmon and ASQ) system. In addition to the spin-conserving transmon transitions (blue and purple), there are two ``mixed'' transition lines involving both spin and transmon degrees of freedom in the presence of coherent coupling between them \cite{Pita-Vidal:2022} (dashed and dotted). The pure spin-flip transitions (orange and yellow) reflect the different spin-orbit splitting, which is here found to be much stronger for the $i=0$ transmon ground state (yellow) than the $i=1$ transmon excited state (orange). 
The general parameter dependence of the amplitude of SOC splitting in the different $i$ levels of the transmon ASQs is an interesting open question for future work.

\section{Applications}
\label{sec:outlook}

Having established that the presented model reproduces all key features of the QD physics, as well as phenomena arising from the interplay of the Josephson effect and charging,
we now show that the method is able to address more challenging problems such as those of strong-coupling between the QD and transmon degrees of freedom, time-dependent problems, as well as the calculation of matrix elements that quantify the possibility of driving various transitions.

\subsection{Strong coupling of transmon and spin degrees of freedom}
\label{strong}

In the absence of SOC $S_z$ is a conserved quantum number, so with increasing magnetic field the $\ket{i=0, \uparrow}$ and $\ket{i=1, \downarrow}$ states cross at a certain field strength. In the presence of the SOC, however, the states instead mix which leads to an avoided crossing. 
This effect can be utilized to implement coupling between two qubits, one encoded in the first two transmon states and the other in the spin of the QD electron, see Ref.~\onlinecite{Pita-Vidal:2022} for an experimental realization. 

\begin{figure*}
    \centering
    \includegraphics[width=1. \textwidth]{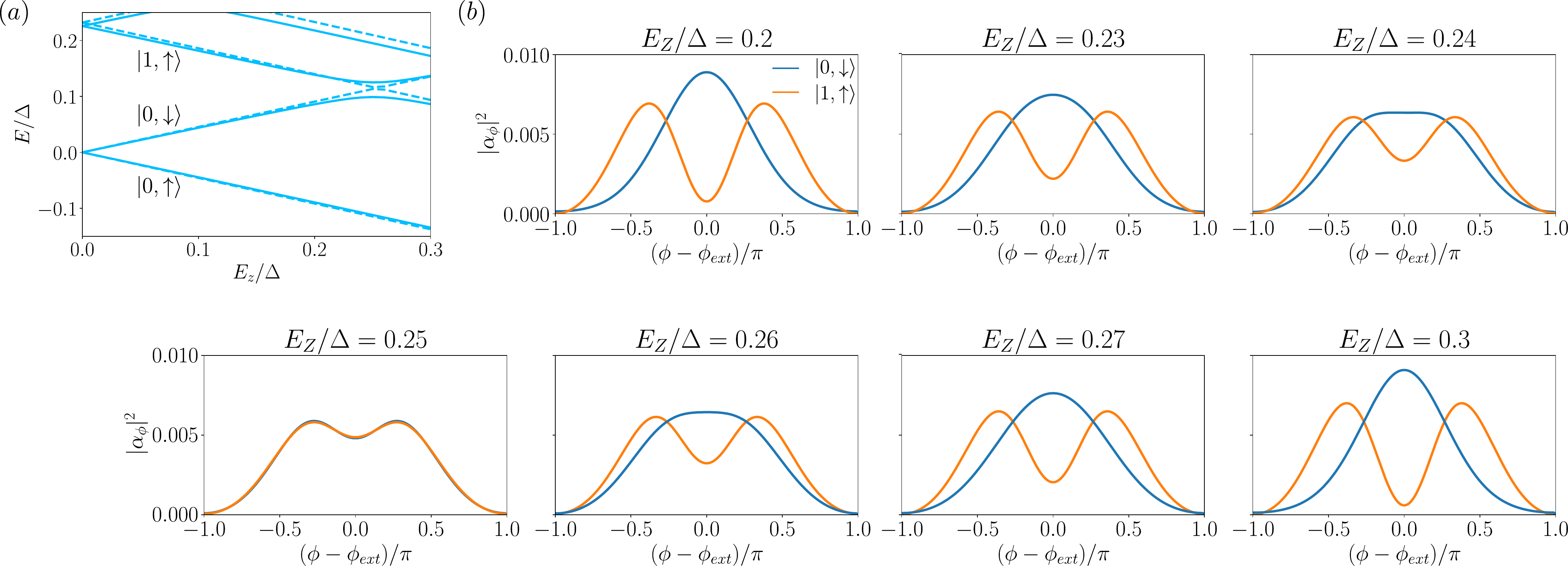}
    \caption{SOC-induced level repulsion.
    \textbf{(a)} 
    Spectra of the doublet manifold with increasing Zeeman splitting of the QD level with (dashed) no spin-orbit coupling and with (solid) $t_{sc} = 0.2\Delta$, $v_{\uparrow\downarrow} = 0.2\Delta$. In both cases, the ground state at $E_Z=0$ is set to zero energy.
    \textbf{(b)} 
    Amplitudes in the $\phi$-space of the $\ket{0, \uparrow}$ and $\ket{1, \downarrow}$ eigenstates through the avoided crossing.    
    Other parameters: $U/\Delta=3$, $\epsilon = -\frac{U}{2}$, $v/\Delta=0.5$, $t_p/\Delta = 0.1$, $\phiext = \pi$, producing $\EJeff \approx 0.22 \Delta$. $\Ec = 0.02\Delta \approx 0.1 \EJeff$. $N=601$ and $n_0 = 300$.
    }
    \label{fig:magnetic_mixing}
\end{figure*}

Here we present a minimal example of such strong mixing effects. Fig.~\ref{fig:magnetic_mixing}(a) shows the splitting of the doublet states with increasing Zeeman splitting on the QD, $E_Z$.
Dashed lines correspond to the case with no SOC, while solid lines exhibit the avoided crossing for finite SOC. 
Our method enables direct access to the wavefunction, which in general is not a product state of the two subsystems, i.e., the two degrees of freedom become entangled.
In Fig.~\ref{fig:magnetic_mixing}(b) we plot $\vert \alpha_\phi \vert^2$ for the two states for a range of Zeeman energies $E_Z$ through the avoided crossing, gauging the mixing in the transmon subspace.
The $\ket{0, \uparrow}$ state evolves from a single-peak shape, characteristic of the transmon ground state, to two peaks at the crossing point $E_Z/\Delta = 0.25$. The single peak is reinstated as the energy difference between the states increases at large $E_Z$.
We find that this behavior is ubiquitous and largely independent of the values of the parameters.

Such avoided crossing has been recently observed experimentally in an ASQ-transmon device, see Fig.~4(c,d) in Ref.~\onlinecite{Pita-Vidal:2022}.
We return to this phenomenon in Sec.~\ref{matel}.

\subsection{Exact time evolution}
\label{time}

Because of the exponential reduction of the Hilbert space achieved with the flat-band approximation, by obtaining the full eigendecomposition of the Hamiltonian via exact diagonalisation it is straightforward to compute the time evolution and calculate various non-equilibrium properties by numerically integrating the time-dependent Schr\"{o}dinger equation.
Such calculations are out of reach of other methods. 

\begin{figure*}
    \centering
    \includegraphics[width=0.7 \textwidth]{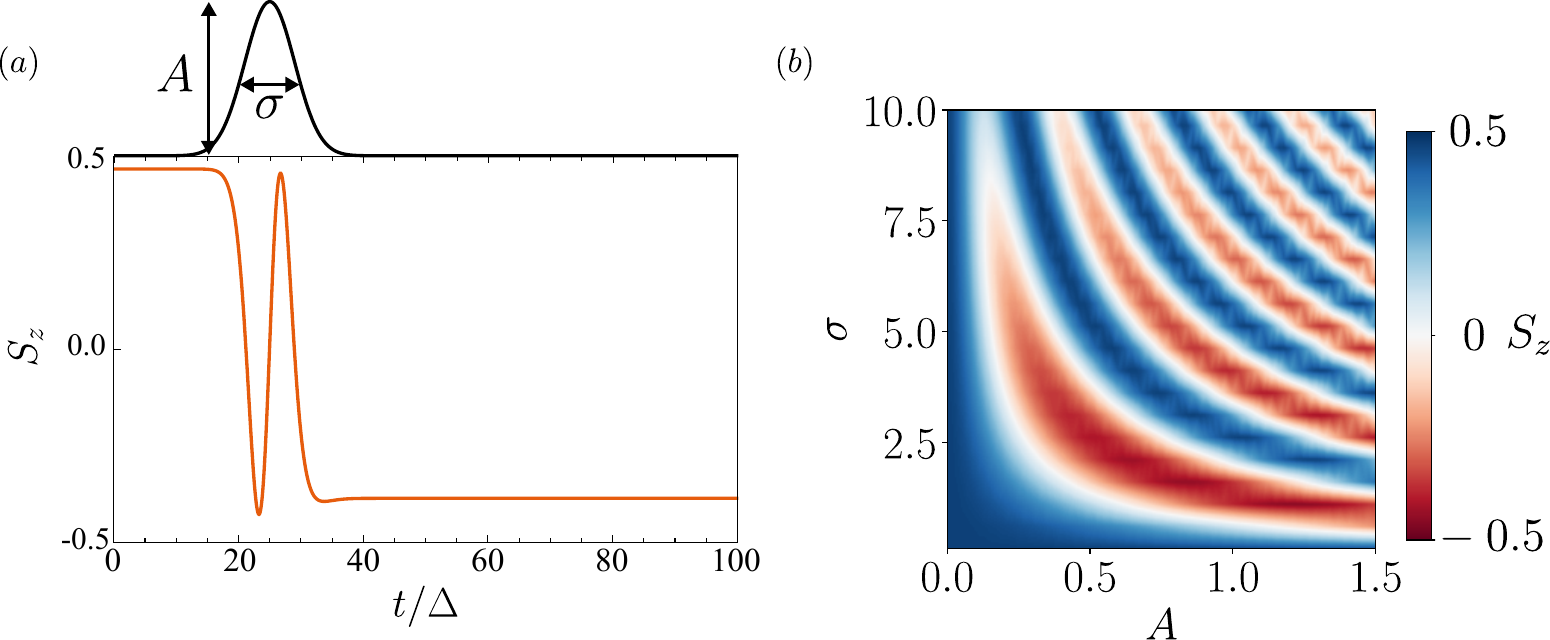}
    \caption{
    Coherent evolution in response to spin-rotating pulses.
    \textbf{(a)} 
    Time evolution of the expectation value of $S_z$ of the QD spin when applying a Gaussian $\hat{S}_x$ pulse (shown on top).
    \textbf{(b)} 
    Expectation value of $S_z$ at long times, $t \gg \sigma$, for a pulse of height $A$ and width $\sigma$.
    Other parameters: $U=3\Delta$, $\epsilon = -\frac{U}{2}$, $v/\Delta=0.5$, $E_Z/\Delta = 0.1$, $t_p = 0$, $\Ec = 0$, no SOC.
    }
    \label{fig:Sx_pulse}
\end{figure*}

A minimal example of Rabi oscillations is shown in Fig.~\ref{fig:Sx_pulse} \cite{Vion2002,Vion2003}. 
We turn on a small Zeeman splitting $E_Z/\Delta = 0.1$ to split the two doublet states, and apply a Gaussian spin-flip pulse (for simplicity we assume that we are at resonance, that we work in the interaction picture, and that we have dropped the counter-rotating terms in the rotating-wave approximation \cite{Vion2003}):
\begin{equation}
    H'(t) = A e^{-(t-t_0)^2/2\sigma^2} \hat{S}_x.
\end{equation}
The time evolution of the expectation value of $S_z$ at the QD is shown in Fig.~\ref{fig:Sx_pulse}(a).
For the chosen values of $A$ and $\sigma$ the pulse causes a single oscillation of the spin, which then settles at $S_z \sim -1/2$.
Fig.~\ref{fig:Sx_pulse}(b) shows the final value of $\langle S_z \rangle$ long after the pulse with varying $A$ and $\sigma$. Typical Rabi fringes are observed.

An obvious utility of such calculations is in predicting optimal pulses for controlling the qubits~\cite{Koch2022}.
This could be done by interfacing our code with a quantum optimal control library where the required input is the system Hamiltonian, eg. Ref.~\onlinecite{Rossignolo2023}.
For example, one could predict and optimize two-qubit gates in the ASQ-transmon devices \cite{Pita-Vidal:2022}.

\subsection{Transition matrix elements}
\label{matel}

\begin{figure*}
    \centering
    \includegraphics[width=2.\columnwidth]{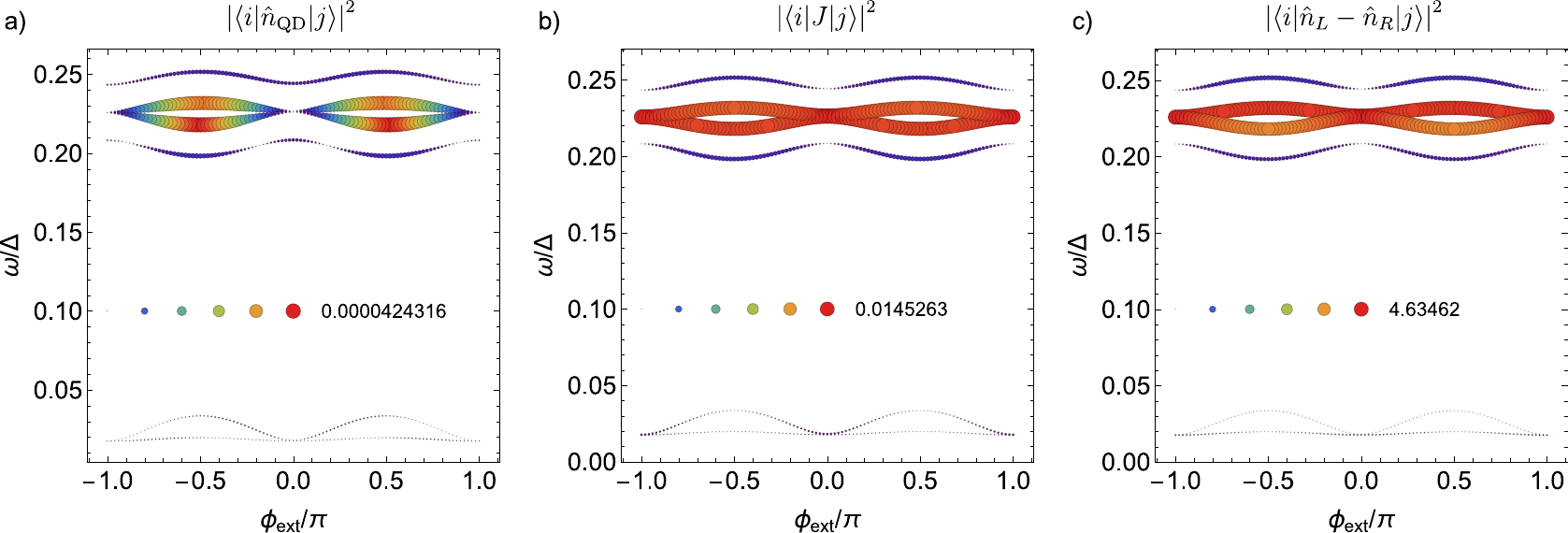}
    \caption{
    Matrix elements for transitions: overview of the $\phiext$ dependence. 
    a) Charge operator, $|\langle i | \hat{n}_\mathrm{QD} | j \rangle|^2$, 
    b) current operator, $|\langle i | J | j \rangle|^2$, with $J=\partial\hat{H}/\partial \phi_\mathrm{ext}$. 
    c) dipole operator, $|\langle i | \hat{n}_\mathrm{L}-\hat{n}_\mathrm{R} | j \rangle|^2$.
    Vertical positions indicate the transition frequency.
    Symbol areas are proportional to the squared
    absolute values of the matrix elements; the number indicated in the legend corresponds to the maximum value attained (the two charge operators are dimensionless, while the ``current'' operator is given in the energy units of $\Delta$). The numbers are not directly comparable, because the overall transition probability also depends on different prefactors.
    Parameters: $U/\Delta=3$, $\epsilon = -\frac{U}{2}$, $v/\Delta=0.5$, $t_{sc} = 0.2\Delta$, $v_{\uparrow\downarrow} = 0.2\Delta$, $t_p/\Delta = 0.1$, $\phiext = \pi$, producing $\EJeff \approx 0.22 \Delta$. $\Ec = 0.02\Delta \approx 0.1 \EJeff$ (as in Fig.~\ref{fig:magnetic_mixing}), for a perpendicular magnetic field of $E_Z/\Delta=0.02$. $N=101$ and $n_0 = 50$.
    }
    \label{fig:matel1}
\end{figure*}

\begin{figure*}
    \centering
    \includegraphics[width=2\columnwidth]{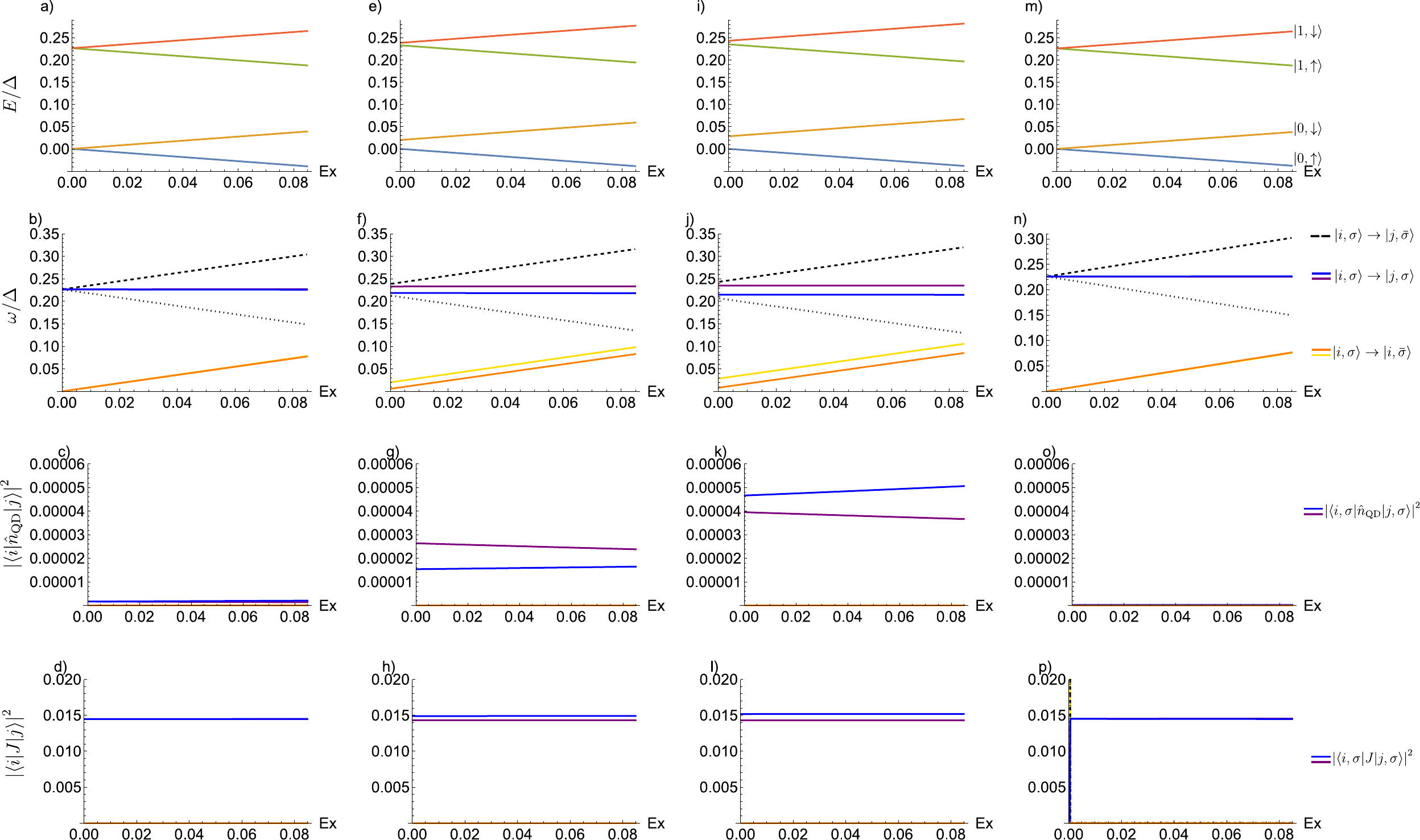}
    \caption{
    Magnetic field dependence: parallel field $E_x$.
      First column: $\phiext=0$, second column: $\phiext=\pi/4$, third column: $\phiext=\pi/2$, last column: $\phiext=\pi$.
    a,e,i,m) Eigenenergies. 
    b,f,j,n) Transition frequencies. 
    c,g,k,o) Matrix elements for the charge operator.
    d,h,l,p) Matrix elements for the current operator.
    Color scheme for transitions as in Fig.~\ref{fig:splitting}.
        Parameters: $U/\Delta=3$, $\epsilon = -\frac{U}{2}$, $v/\Delta=0.5$, $t_{sc} = 0.2\Delta$, $v_{\uparrow\downarrow} = 0.2\Delta$, $t_p/\Delta = 0.1$, $\Ec = 0.02\Delta$. $N=101$ and $n_0 = 50$.
    }
    \label{fig:matel2}
\end{figure*}

\begin{figure*}
    \centering
    \includegraphics[width=2\columnwidth]{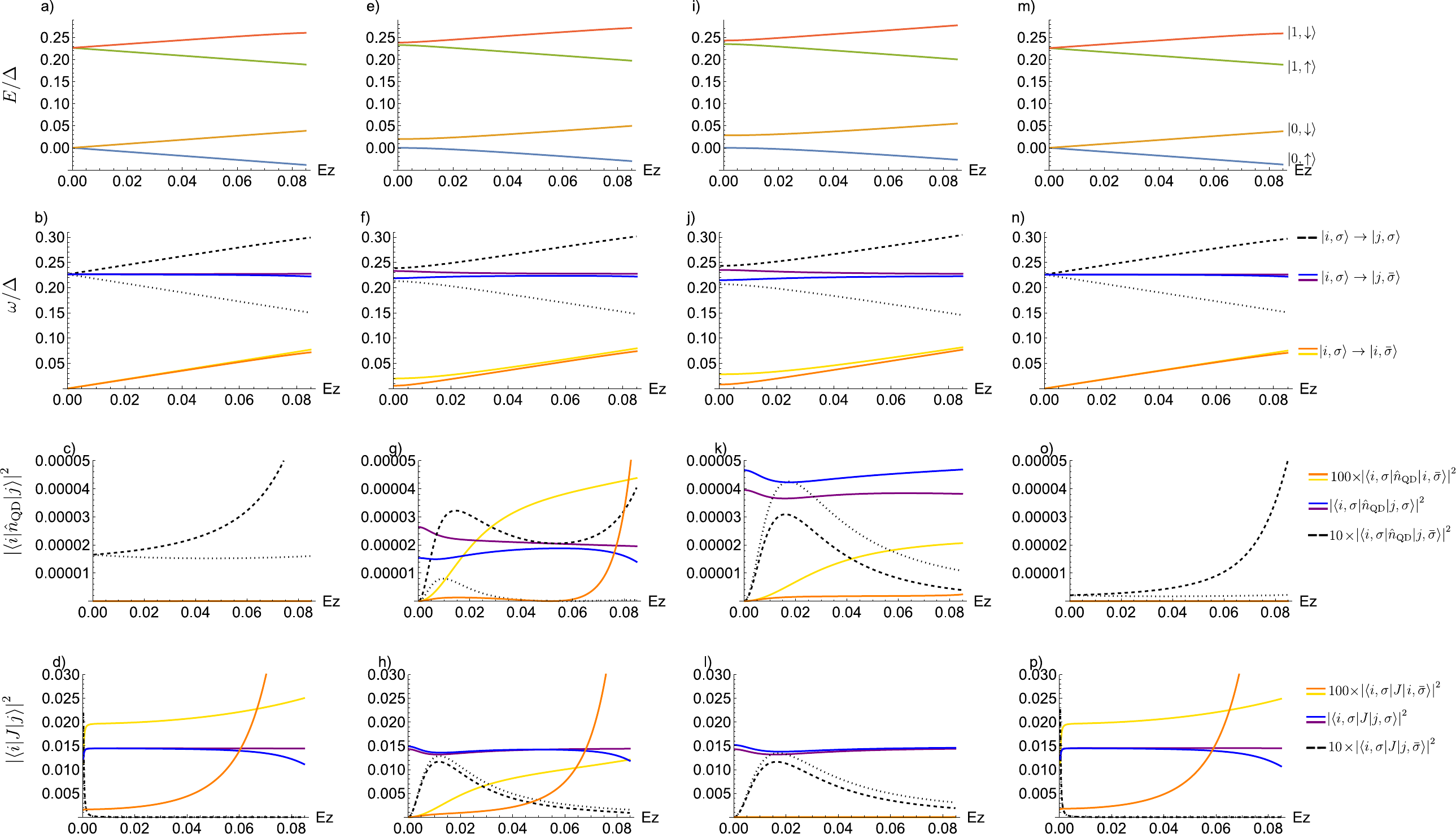}
    \caption{
    Magnetic field dependence: perpendicular field $E_z$ (indicated in units of $\Delta$). 
    For easier comparison of matrix elements for different types of transitions on the same vertical scale, we multiply all matrix elements for pure spin-flip transition matrix elements (yellow, orange) by 100, and all matrix elements for mixed transitions (dashed, dotted) by 10; the matrix elements for pure transmon transitions (blue, purple) are not scaled.
    At the upper range of magnetic field we start to observe anomalies due to the anticrossing between the
    $\ket{1,\downarrow}$ and $\ket{2,\uparrow}$ states that occurs at $E_z/\Delta \approx 0.095$.
       Parameters as in Fig.~\ref{fig:matel2}.
    }
    \label{fig:matel3}
\end{figure*}

\begin{figure}
    \centering
    \includegraphics[width=0.7\columnwidth]{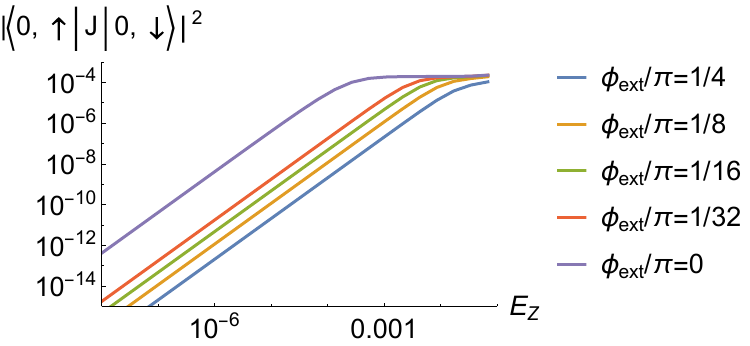}
    \caption{
    Asymptotic low-field behavior of current matrix elements for spin-flip transitions at $\phiext \approx 0$.
    Parameters as in Fig.~\ref{fig:matel3}.
    }
    \label{fig:matelsfn}
\end{figure}

The Hamiltonian includes an effective low-energy description of the QD and its coupling to the rest of the circuit, at the level of single-orbital Anderson impurity model with extensions for the SOC.
While approximate, this description is nonetheless microscopically realistic, it contains all terms which are believed to be relevant (in the renormalisation group sense, as well as considering the symmetry constraints), and it is most likely adequate to address many questions about the interplay between the different degrees of freedom. 
In particular, one can investigate various coupling mechanisms that permit the qubit control. 
In the experiments of Ref.~\cite{Pita-Vidal:2022}, the transitions were driven by applying microwave pulses to the QD gate electrode, which has a dominant capacitive coupling to the QD orbital. All types of transitions shown in Fig.~\ref{fig:splitting} were observed, both pure transmon, pure spin-flip, as well as mixed transitions. 
Due to the SOC, the electric field modulates not only the orbital wavefunction in real space but indirectly also the spin degrees of freedom, a phenomenon known as the electric dipole spin resonance (EDSR) \cite{Golovach2006,Nowack2007,PioroLadrire2008,NadjPerge2010,vandenBerg2013,Pita-Vidal:2022}. 
To shed light on the physical mechanisms of state transitions we now investigate the off-diagonal matrix elements of some key operators, in particular the charge operator, the electric dipole $\sum q_i \mathbf{r}_i$, and the current operator. 
For simplicity, we use the following operating definitions: the charge operator is taken to be the QD occupancy operator $n_\mathrm{QD}$, the dipole is defined as $n_L-n_R$ (which does not involve the QD directly, but contains information about the charge asymmetry in the device), and the current is defined as $J=\partial H/\partial \phiext$.
The current matrix element is calculated using the off-diagonal form of the Hellmann-Feynman theorem:
\begin{equation}
J_{ij} = \langle i | \frac{\partial H}{\partial \phiext} | j \rangle
= (E_i-E_j) \left( \frac{\partial \langle i|}{\partial \phiext} \right) | j \rangle,
\end{equation}
where $E_i$, $E_j$ are the eigenenergies of the states $\ket{i}$ and $\ket{j}$. To calculate the derivative we use finite differences, taking $\Delta\phiext=0.01\pi$.

We first consider the $\phi$-dependence of these operators for an experimentally realistic parameter set, Fig.~\ref{fig:matel1}, at finite magnetic field applied in the direction that is perpendicular to the SOC spin-polarization direction (``perpendicular field''). For all three operators considered, we generally observe that the matrix elements are the largest for the transmon transitions, the weakest for the pure spin-flip transitions, and intermediate for the mixed transitions. 
The operators considered couple to the spin degree of freedom only indirectly, via the SOC, which naturally explains this hierarchy.
Importantly, we also observe different flux dependencies, depending on the operator considered and the transition type.
The matrix elements for the charge operator, Fig.~\ref{fig:matel1}a), have a nearly sinusoidal dependence for the pure transmon transition; they reach a maximum value at $\phiext \approx \pm \pi/2$, and they go through a zero at $\phiext=0$ and $\phiext=\pm \pi$. 
The pure spin-flip transitions are anharmonic, with a maximum close to $\phiext=\pm \pi/4$, and go through a zero at $\phiext=0$ and $\phiext=\pm \pi$. The mixed transitions have a maximum at $\phiext \approx \pm \pi/3$ and remain finite at both $\phiext=0$ and $\phiext=\pm \pi$.
The matrix elements for the current operator, Fig.~\ref{fig:matel1}b), have a weak $\phiext$ dependence for the transmon transitions. Spin-flip and mixed transitions are strongly $\phiext$-dependent and they are out-of-phase (not easily seen in Fig.~\ref{fig:matel1} due to the chosen scale): the pure spin-flip matrix elements are maximal at $\phiext=0$ and go through zero for $\phiext=\pi/2$; the mixed transition matrix element are, conversely, large at $\phiext=\pi/2$ and go through zero at $\phiext=0$.
The matrix elements for the dipole moment, Fig.~\ref{fig:matel1}c), have a similar $\phi$ dependence as those for the current, although they are not exactly proportional. Because of this similarity, in the following we only consider the charge and current operators, assuming that these two operators represent the two important classes of the $\phiext$ dependence. (We have also considered the quadrupole operator with an operational definition $n_\mathrm{L}+n_\mathrm{R}$; in our model it brings no new information because it is trivially related to the charge operator $n_\mathrm{QD}$.)

The experiments suggest that driving becomes easier with the application of the magnetic field \cite{Pita-Vidal:2022}. 
A magnetic field that is parallel to the SOC spin-polarization direction (``parallel field'') has little effect, see Fig.~\ref{fig:matel2}. The most striking feature is the fact that only transmon transitions show finite matrix elements in this case, spin-flip and mixed transitions are altogether absent. From this we immediately conclude that within our model a perpendicular component of magnetic field is necessary to enable transitions involving the spin degree of freedom if the coupling is only via charge or current.

We next investigate the case of a perpendicular magnetic field, Fig.~\ref{fig:matel3}. 
The pure transmon transition matrix elements (blue, violet) are not affected strongly by the field, yet they do show some non-trivial field dependence. The pure spin-flip and mixed transitions have even more complex field variation which furthermore depends on the value of $\phiext$. The pure spin-flip matrix elements (orange, yellow) remain zero for the charge operator at $\phiext=0$, and for the current operator at $\phiext=\pi/2$; for general $\phiext$, they increase with the field, although there is a sizable dependence on the transmon level. Mixed transitions (dashes, dotted) are found to have non-monotonic field dependence in most cases, with a maximum value attained for some intermediate field strength. We note that the cases of $\phiext\approx 0$ and $\phiext \approx \pi$ appear to be anomalous, and the $E_z\to0$ dependence of current matrix elements is misleading: at very low field values the matrix elements for transitions involving spin-flips actually drop to zero rather the saturate, see Fig.~\ref{fig:matelsfn}. Nevertheless, even a small perpendicular magnetic field is apparently sufficient to enable the spin-flip transitions at these $\phiext$ settings.

It turns out that the role of perpendicular field can be explained via a simplified zero-bandwidth-approximation (ZBA) calculation, see App.~\ref{appB}: the low-field behavior is indeed found to be linear in $E_z$. Such calculations also reproduce some trends observed in the $\phi$-dependence (see also App.~\ref{appC}).

\begin{figure}
    \centering
    \includegraphics[width=0.9\columnwidth]{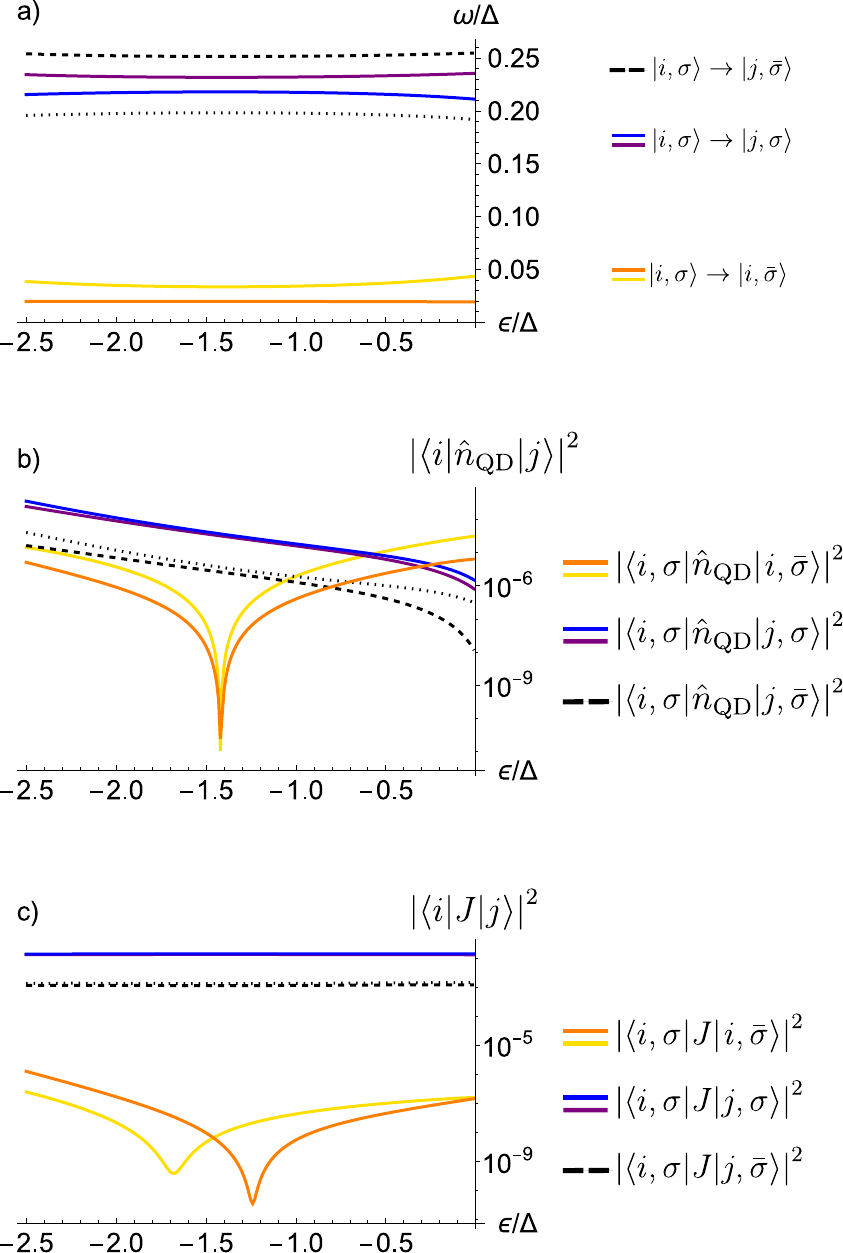}
    \caption{
    Quantum dot level $\epsilon$ dependence. 
    Parameters: $U/\Delta=3$, $v/\Delta=0.5$, $t_{sc} = 0.2\Delta$, $v_{\uparrow\downarrow} = 0.2\Delta$, $t_p/\Delta = 0.1$, $\Ec = 0.02\Delta$, $\phiext=\pi/2$, $E_z/\Delta=0.02$. $N=101$ and $n_0 = 50$.
    }
    \label{fig:matel4}
\end{figure}

\begin{figure}
    \centering
    \includegraphics[width=0.9\columnwidth]{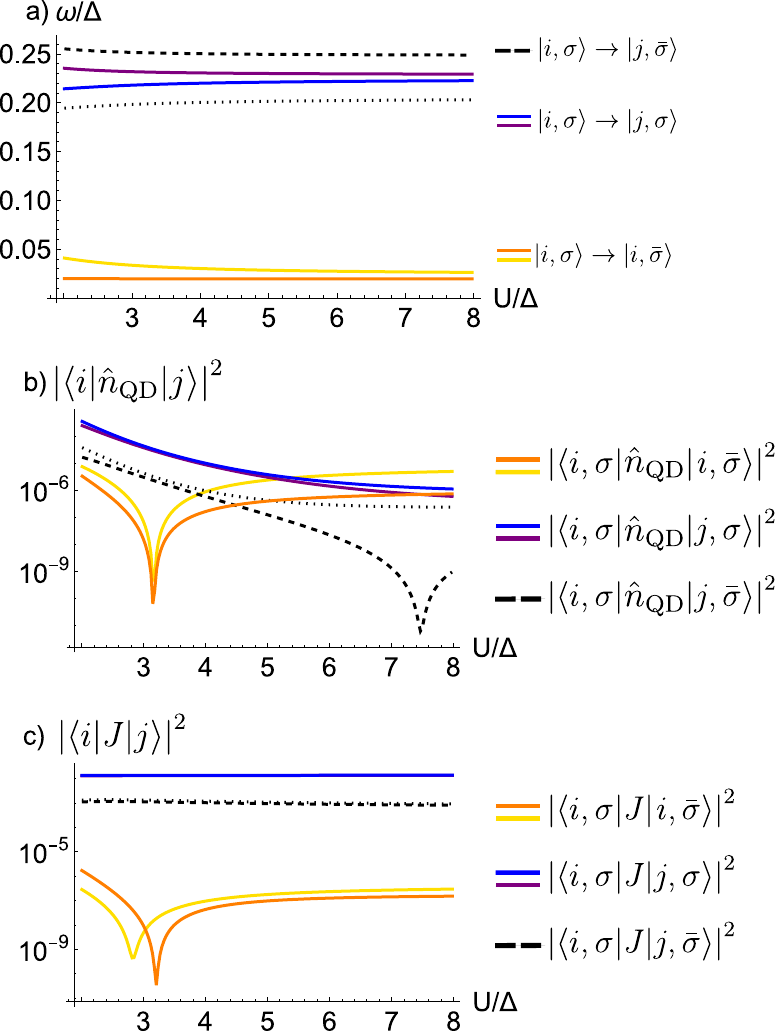}
    \caption{
    Quantum dot electron-electron repulsion $U$ dependence. 
    Parameters: $\epsilon/\Delta=-1.5$, $v/\Delta=0.5$, $t_{sc} = 0.2\Delta$, $v_{\uparrow\downarrow} = 0.2\Delta$, $t_p/\Delta = 0.1$, $E_c/\Delta=0.02$, $\phiext=\pi/2$, $E_z/\Delta=0.02$.
    }
    \label{fig:matel7}
\end{figure}

For completeness, we now study the variation of eigenenergies and matrix elements upon variation of other key model parameters. 
We consider the case of small perpendicular field $E_z=0.02\Delta$ at $\phiext=\pi/2$.
The variation of the QD level $\epsilon$, see Fig.~\ref{fig:matel4}, has a weak effect on the energies, but we observe a large effect on the charge matrix elements, which is monotonic for transmon and mixed transitions, while for pure spin-flip transitions it changes sign not far from the particle-hole symmetric point at $\epsilon=-U/2=-1.5\Delta$. This behavior is also predicted by the ZBA calculation in App.~\ref{appB}. Furthermore, we note that the relative magnitude of pure spin-flip and mixed transition matrix elements is a strong function of $\epsilon$. For the current operators, the dependence on $\epsilon$ is very weak for the transmon and mixed transitions; for the pure spin-flip transition there is again a sign change, but these matrix elements are anyhow quite low. Rather similar results are obtained upon variation of the electron-electron repulsion $U$, see Fig.~\ref{fig:matel7}.

\begin{figure}
    \centering
    \includegraphics[width=0.9\columnwidth]{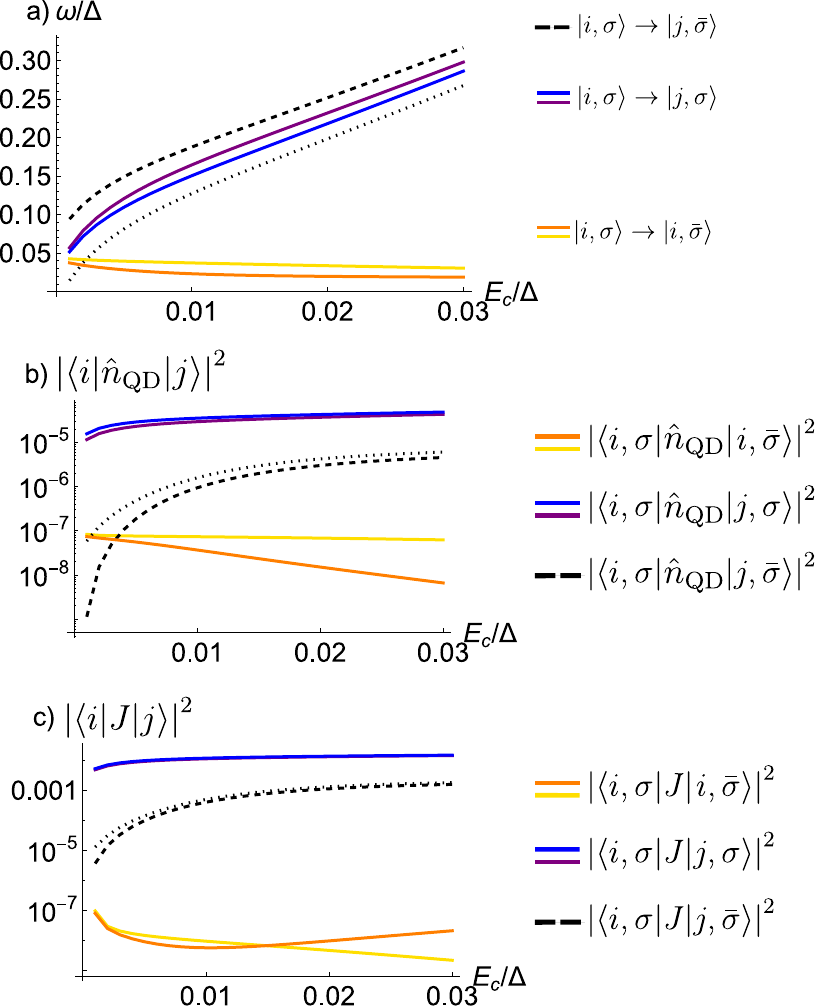}
    \caption{
    Charging energy $E_c$ dependence. 
    Parameters: $U/\Delta=3$, $\epsilon=-U/2$, $v/\Delta=0.5$, $t_{sc} = 0.2\Delta$, $v_{\uparrow\downarrow} = 0.2\Delta$, $t_p/\Delta = 0.1$, $\phiext=\pi/2$, $E_z/\Delta=0.02$. $N=101$ and $n_0 = 50$.
    }
    \label{fig:matel5}
\end{figure}

The charging energy $E_c$, see Fig.~\ref{fig:matel5}, strongly affects the transmon transition frequency, as expected. More interestingly, increasing $E_c$ suppresses the pure spin-flip frequency, i.e., it renormalizes the SOC splitting, the effect being particularly strong for the excited $i=1$ level. 
For both charge and current matrix elements, the strongest $E_c$ dependence is observed in the mixed transitions, while pure transmon and pure spin-flip transitions (esp. for $i=0$) appear to be less sensitive to the value of $E_c$.
We note that the proposed model is the minimal Hamiltonian that allows to study how the QD properties are renormalized by the phase fluctuations in the superconducting circuit. One could reexamine, for example, the impurity Knight shift explored in Ref.~\onlinecite{knight} and determine how a finite $E_c$ affects the renormalization of the Zeeman splitting by the coupling to the superconducting leads.

\begin{figure}
    \centering
    \includegraphics[width=0.9\columnwidth]{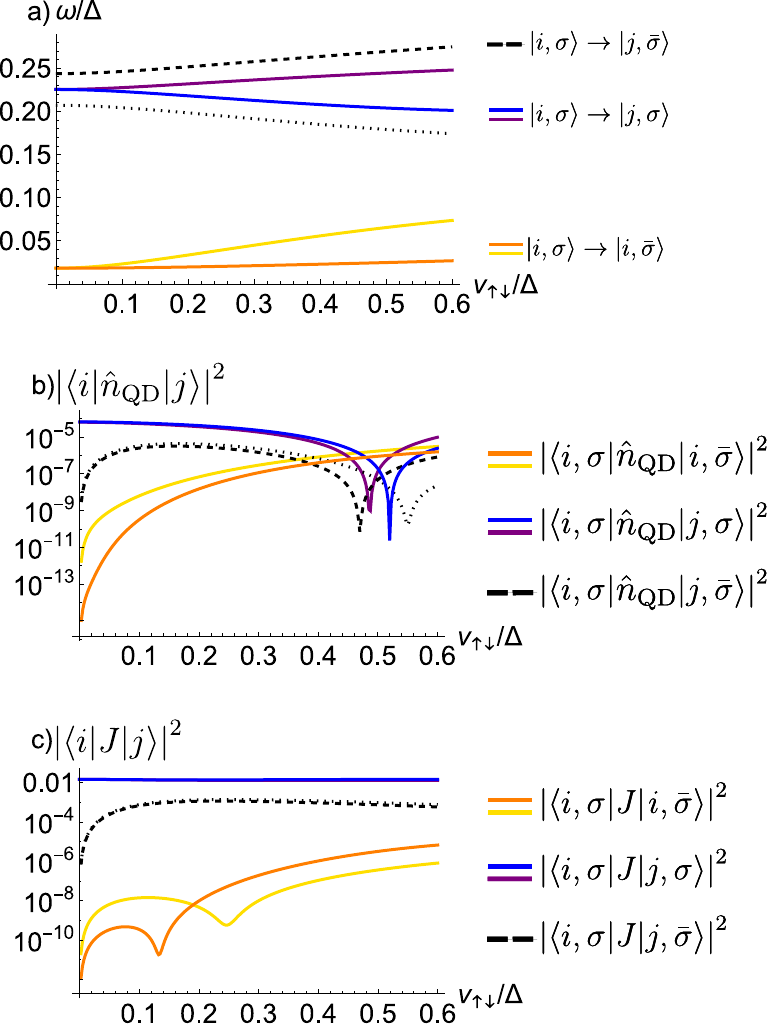}
    \caption{
    SOC (spin-flip hybridization $v_{\uparrow\downarrow}$) dependence. 
    Parameters: $U/\Delta=3$, $\epsilon=-U/2$, $v/\Delta=0.5$, $t_{sc} = 0.2\Delta$, $t_p/\Delta = 0.1$, $E_c/\Delta=0.02$, $\phiext=\pi/2$, $E_z/\Delta=0.02$. $N=101$ and $n_0 = 50$.
    }
    \label{fig:matel6}
\end{figure}

The strength of SOC is proportional to the spin-flip hybridisation $v_{\uparrow\downarrow}$ (in fact, it is proportional to the product of $v v_{\uparrow\downarrow} t_\mathrm{sc}$, see Ref.~\cite{Bargerbos:22b}). Indeed, the SOC level splittings are directly reflected in the approximately linear trends of transition frequencies, see Fig.~\ref{fig:matel6}. As a general trend (with some exceptions for weak SOC), the pure spin-flip matrix elements tend to increase with $v_{\uparrow\downarrow}$, while the mixed transition tend to decrease; this is the case for both charge and current operators. The transitions involving the transmon degree of freedom show a sign change of matrix elements for the charge operator at some large value of SOC (this is also reflected in mixed transitions). As for the spin-flip transitions, we observe a sign change for the current operator. This is, in fact, predicted by the ZBA calculation, see App.~\ref{appB}, which predict a proportionality of transition matrix elements to $v^2-v_{\uparrow\downarrow}^2$. Clearly, this is renormalized by the coupling to transmon degrees of freedom, so that the cancellation point becomes transmon level dependent.

\begin{figure}
    \centering
    \includegraphics[width=1.\columnwidth]{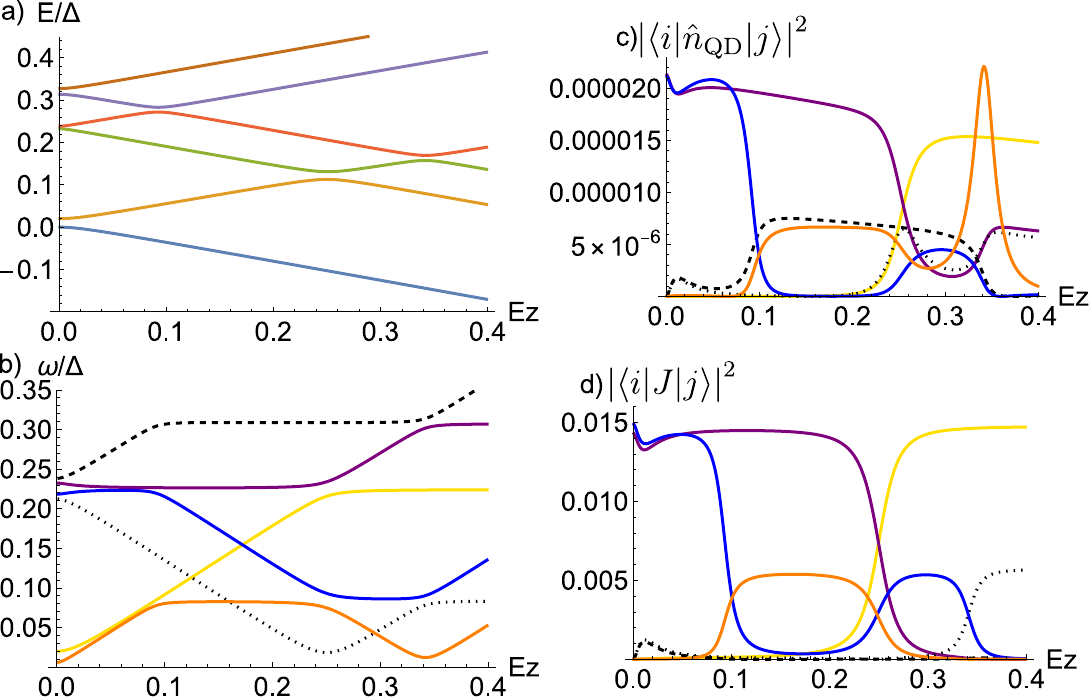}
    \caption{
        Perpendicular magnetic field dependence. Parameters as in Fig.~\ref{fig:matel3}, except $\phiext=(3/4)\pi$. The values of matrix elements are not rescaled in any way.
    }
    \label{fig:matel9}
\end{figure}

We now return to the case of strong ASQ-transmon coupling, already introduced in Sec.~\ref{strong}. We thus extend the plots similar to that in Fig.~\ref{fig:matel3} to a wider range of magnetic fields to fully reveal the variation of the transition matrix elements in the vicinity of anticrossings, of which there are three in the range considered: between $\ket{1,\downarrow}$ and $\ket{2,\uparrow}$ at $E_z/\Delta \approx 0.095$, between $\ket{0,\downarrow}$ and $\ket{1,\uparrow}$ at $E_z/\Delta \approx 0.25$ (the case discussed in relation to Fig.~\ref{fig:matel9} and the one experimentally explored in Fig.~3c,d) in Ref.~\onlinecite{Pita-Vidal:2022}), and between $\ket{1,\downarrow}$ and $\ket{1,\uparrow}$ at $E_z/\Delta \approx 0.34$. It should be noted that we here use labelling that adiabatically continues the zero-field labels, although it is clear, for example, that the third anticrossing occurs between the states that actually correspond to the $i=2$ and $i=0$ transmon levels. Let us focus on the anticrossing at $E_z/\Delta \approx 0.25$. The anticrossing between $\ket{0,\downarrow}$ and $\ket{1,\uparrow}$ (panel a) is indirectly revealed through the behavior of transition frequencies as an ``anticrossing'' between a pure transmon transition and a mixed (spin-flipping transmon) transition (yellow and purple lines involving the $\ket{0,\uparrow}$ ground state in panel b, but also orange and blue lines involving the $\ket{1,\downarrow}$ excited state). Across the anticrossing range the matrix elements are changing between two asymptotic value, which is a signature of the changing nature of the states. Importantly, at the point of closest approach all matrix elements are sizable, even if the asymptotic value on one side of the transition is very small. 

The examples in this subsection clearly illustrate that the proposed method can provide guidance in choosing system parameters that lead to the optimal operating point, both in terms of level energies as well as for the transition matrix elements. We have observed that the variation of the matrix elements is in general not trivial, it can easily span multiple decades in amplitude, it is often non-monotonic, and there are sign changes. While some general trends for pure spin-flip and pure transmon transition can be extracted for simplified ZBA calculation that do not involve the dynamical superconductor (transmon) degrees of freedom (App.~\ref{appB} and \ref{appC}), the full model is required to capture all the details and to account for the mixed transitions.

\section{Conclusion}

We have investigated a system of a QD embedded in a junction between two superconductors from the viewpoint of the charge-conserving Richardson model of superconductivity. 
We argue that this approach enables clearer understanding of various properties of the system, especially those related to the emergence of the superconducting phase difference and its dynamics.
Most importantly, it is able to access parameters regimes that BCS-type approximations cannot, most notably the case of finite charging energy in the superconducting islands.
The model treats on the same level all important physical processes---the impurity physics of the interacting QD, the charging energy and pairing terms in the superconductors, and the Josephson effect. Furthermore, by including spin-orbit coupling this becomes a model for Andreev spin qubits in transmon devices.

The central step of our solution is the flat-band approximation, which consists of neglecting the kinetic energy in the superconductors.
This leads to an exponential reduction of the Hilbert space while retaining all states through which the QD is coupled to the SCs and are important for the low-energy physics of the system.

The obtained Hamiltonian is similar, in spirit, to the zero-bandwidth BCS approximation for the superconducting Anderson impurity model, however it retains  information about the distribution of the Cooper pairs between the two superconductors, as in the conventional transmon Hamiltonian, thereby combining aspects of both approaches.
It is formulated in the conjugate space of well-defined charge, which enables natural implementation of the charging energy terms in the SCs and allows us to capture the quantum fluctuations of the phase difference.

We have demonstrated that the model reproduces all features expected of a quantum dot Josephson junction in the limit of zero charging energy, and we have singled out the parameter regimes where standard approximations fail, pointing out the importance of quasiparticles in the superconductor induced by the pair-breaking effects of the ``magnetic'' quantum dot.
Next, we investigated the effect of the charging energy and its interplay with the Josephson effect. 
We observed and quantified the evolution between the limits of well-defined phase difference and well-defined charge difference on the level of eigenstate properties and quantum fluctuations of phase. Finally, we showcased the applications of the method with three experimentally relevant problems: 1) the case where the transmon and the QD degrees of freedom mix, 2) the calculation of QD spin flipping induced by a pulse, 3) the investigation of transition matrix elements.

The model can be applied to other states as well. While in this work we have focused on the doublet subspace, as appropriate for Andreev spin qubits, it is equally possible to study the singlet subspace which is relevant for Andreev level qubits based on the superpositions of empty and doubly occupied quantum dot orbital.

Finally, noting that the Hilbert space is only moderately large, it seems quite feasible to include relaxation and dephasing effects in the model through appropriate Lindblad operators and to study decoherence in the associated open quantum system.

\begin{acknowledgments}
We thank Marta Pita Vidal and Arno Bargerbos for suggestions concerning the transition matrix elements and other applications of the presented method. We acknowledge the support of the Slovenian Research and Innovation Agency (ARIS) under P1-0416 and J1-3008.
\end{acknowledgments}

\clearpage 
\appendix

\section{Basis states and Hamiltonian generation}
\label{appA}

Here we list the basis states that span the singlet ($S=0$, $S_z = 0$) and the doublet ($S=1/2$, $S_z = +1/2$) subspaces, and describe how the Hamiltonian matrix is generated.

The code, including the matrices, is available on GitHub \cite{flat2023}.

\subsection{Singlet states}

The singlet manifold has dimension 14, consisting of one state with no unpaired particles:
\begin{itemize}
    \item $ \ket{m_L, 0, m_R, 0} $,
\end{itemize}
six states with two unpaired particles:
\begin{itemize}
    \item $ \ket{m_L, 0, m_R, 2} $,
    \item $ \ket{m_L, 2, m_R, 0} $,
    \item $ d^\dagger_\downarrow d^\dagger_\uparrow \ket{m_L, 0, m_R, 0} $,
    \item $ \left( d^\dagger_\uparrow \ket{m_L, \downarrow, m_R, 0} - d^\dagger_\downarrow \ket{m_L, \uparrow, m_R, 0} \right)/\sqrt{2}$,
    \item $ \left( d^\dagger_\uparrow \ket{m_L, 0, m_R, \downarrow} - d^\dagger_\downarrow \ket{m_L, 0, m_R, \uparrow} \right)/\sqrt{2}$,
    \item $ \left( \ket{m_L, \uparrow, m_R, \downarrow} - \ket{m_L, \downarrow, m_R, \uparrow} \right)/\sqrt{2}$,
\end{itemize}
six states with four unpaired particles:
\begin{itemize}
    \item $ \ket{m_L, 2, m_R, 2} $,
    \item $ d^\dagger_\downarrow d^\dagger_\uparrow \ket{m_L, 2, m_R, 0} $,
    \item $ d^\dagger_\downarrow d^\dagger_\uparrow \ket{m_L, 0, m_R, 2} $,

    \item $ \left( d^\dagger_\uparrow \ket{m_L, \downarrow, m_R, 2} - d^\dagger_\downarrow \ket{m_L, \uparrow, m_R, 2} \right)/\sqrt{2}$,
    \item $ \left( d^\dagger_\uparrow \ket{m_L, 2, m_R, \downarrow} - d^\dagger_\downarrow \ket{m_L, 2, m_R, \uparrow} \right)/\sqrt{2}$,
    \item $ \left( d^\dagger_\downarrow d^\dagger_\uparrow \ket{m_L, \uparrow, m_R, \downarrow} - d^\dagger_\downarrow d^\dagger_\uparrow  \ket{m_L, \downarrow, m_R, \uparrow} \right)/\sqrt{2}$,
\end{itemize}
and one state with six unpaired particles:
\begin{itemize}
    \item $ d^\dagger_\downarrow d^\dagger_\uparrow \ket{m_L, 2, m_R, 2} $.
\end{itemize}

\subsection{Doublet states}

The doublet manifold also has dimension 14. 
It contains three states with one unpaired particle:
\begin{itemize}
    \item $ d^\dagger_\uparrow \ket{m_L, 0, m_R, 0} $,
    \item $ \ket{m_L, \uparrow, m_R, 0} $,
    \item $ \ket{m_L, 0, m_R, \uparrow} $,
\end{itemize}
eight states with three unpaired particles:
\begin{itemize}
    \item $ d^\dagger_\downarrow d^\dagger_\uparrow \ket{m_L, \uparrow, m_R, 0} $,
    \item $ d^\dagger_\downarrow d^\dagger_\uparrow \ket{m_L, 0, m_R, \uparrow} $,
    \item $ d^\dagger_\uparrow \ket{m_L, 2, m_R, 0} $,
    \item $ d^\dagger_\uparrow \ket{m_L, 0, m_R, 2} $,
    \item $ \ket{m_L, \uparrow, m_R, 2} $,
    \item $ \ket{m_L, 2, m_R, \uparrow} $,
    \item $ \left( d^\dagger_\downarrow \ket{m_L, \uparrow, m_R, \uparrow} - d^\dagger_\uparrow \ket{m_L, \downarrow, m_R, \uparrow} \right) /\sqrt{2}$,
    \item $ \Bigl( d^\dagger_\downarrow \ket{m_L, \uparrow, m_R, \uparrow} + d^\dagger_\uparrow \ket{m_L, \downarrow, m_R, \uparrow} \\
    \hspace{2cm} - 2 d^\dagger_\uparrow \ket{m_L, \uparrow, m_R, \downarrow} \Bigr) /\sqrt{6}$,
\end{itemize}
and three states with five unpaired particles:
\begin{itemize}
    \item $ d^\dagger_\downarrow d^\dagger_\uparrow \ket{m_L, 2, m_R, \uparrow} $,
    \item $ d^\dagger_\downarrow d^\dagger_\uparrow \ket{m_L, \uparrow, m_R, 2} $,
    \item $ d^\dagger_\uparrow \ket{m_L, 2, m_R, 2} $.
\end{itemize}

These are the states with $S_z = +1/2$. In cases of broken time-reversal symmetry the $S_z = \pm 1/2$ degeneracy is broken and the basis should be extended with the corresponding $S_z = -1/2$ states.

\subsection{Hamiltonian generation}

The diagonal part of the Hamiltonian is 
\begin{equation}
    \begin{split}
    H_\mathrm{diag} = & \epsilon n_\QD + U n_{\QD \uparrow} n_{\QD \downarrow} + \frac{g}{2} \sum_{\beta \sigma} f^\dagger_{\beta \sigma} f_{\beta \sigma} \\
    &+ \sum_\beta E_c^{\beta} \left( \sum_\sigma f^\dagger_{\beta \sigma} f_{\beta \sigma} + 2m_\beta - n_0 \right)^2 \\
    \end{split}
\end{equation}
and can be read out directly for each basis state.

The off-diagonal part is generated by first calculating analytical expressions for all matrix elements 
\begin{equation}
    \langle d', m'_L, f'_L, m'_R, f'_R \vert H_\mathrm{hyb} \vert d, m_L, f_L, m_R, f_R \rangle.
    \label{eq:general_matrix_element}
\end{equation}
Only states where a single electron is transferred from the QD to a SC are coupled, so the matrix is sparse, with elements of the type 
\begin{equation}
\begin{split}
    &\langle m'_L, \uparrow, m'_R, 0 \vert H_\mathrm{hyb} d^\dagger_\uparrow \vert m_L,0, m_R, 0 \rangle \\ 
    & \hspace{1cm} = v_L \delta_{m_L, m'_L+1} \delta_{m_R, m'_R}.
    \label{eq:hyb}
\end{split}
\end{equation}
The matrices of $H_\mathrm{hyb}$ in the singlet (Eq.~\eqref{eq:singlet_matrix}) and doublet (Eq.~\eqref{eq:doublet_matrix}) are given below. 
To shorten the notation, we denote $\delta^\beta_x = \delta_{m_\beta,m'_{\beta+x}}$.

% regex to find all t_sc terms in the matrices: &[^&]*t_\\mathrm\{sc\} [^&]* & 
% regex to fix the cases where the fraction is too big: -\\frac\{1\}\{2\}([^&]*) replace with \frac{-$1}{2}

\begin{equation}
\rotatebox{90}{$
\begin{bmatrix}
0 & \frac{-v_L \delta_0^L \delta_0^R}{\sqrt{2}} & \frac{-v_R \delta_0^L \delta_0^R}{\sqrt{2}} & \frac{v_L \delta_{1}^L \delta_0^R}{\sqrt{2}} & \frac{v_R \delta_0^L \delta_1^R}{\sqrt{2}} & 0 & 0 & 0 & 0 & 0 & 0 & 0 & 0 & 0 \\
 \frac{-v_L \delta_0^L \delta_0^R}{\sqrt{2}} & 0 & 0 & 0 & 0 & \frac{v_L \delta_{1}^L \delta_0^R}{\sqrt{2}} & 0 & 0 & 0 & \frac{- v_R \delta_0^L \delta_1^R }{2}& \frac{- \sqrt{3} v_R \delta_0^L \delta_1^R }{2}& 0 & 0 & 0 \\
 \frac{-v_R \delta_0^L \delta_0^R}{\sqrt{2}} & 0 & 0 & 0 & 0 & 0 & 0 & 0 & \frac{v_R \delta_0^L \delta_1^R}{\sqrt{2}} & v_L \delta_{1}^L \delta_0^R & 0 & 0 & 0 & 0 \\
 \frac{v_L \delta_{-1}^L \delta_0^R}{\sqrt{2}} & 0 & 0 & 0 & 0 & \frac{v_L \delta_0^L \delta_0^R}{\sqrt{2}} & 0 & 0 & 0 & \frac{- v_R \delta_0^L \delta_0^R }{2}& \frac{- \sqrt{3} v_R \delta_0^L \delta_0^R }{2}& 0 & 0 & 0 \\
 \frac{v_R \delta_0^L \delta_{-1}^R}{\sqrt{2}} & 0 & 0 & 0 & 0 & 0 & 0 & 0 & \frac{v_R \delta_0^L\delta_0^R}{\sqrt{2}} & v_L \delta_0^L \delta_0^R & 0 & 0 & 0 & 0 \\
 0 & \frac{v_L \delta_{-1}^L \delta_0^R}{\sqrt{2}} & 0 & \frac{v_L \delta_0^L \delta_0^R}{\sqrt{2}} & 0 & 0 & \frac{-v_R \delta_0^L \delta_0^R}{\sqrt{2}} & 0 & 0 & 0 & 0 & \frac{v_R \delta_0^L \delta_1^R}{\sqrt{2}} & 0 & 0 \\
 0 & 0 & 0 & 0 & 0 & \frac{-v_R \delta_0^L \delta_0^R}{\sqrt{2}} & 0 & 0 & 0 & v_L \delta_0^L \delta_0^R & 0 & 0 & 0 & \frac{v_R \delta_0^L \delta_1^R}{\sqrt{2}} \\
 0 & 0 & 0 & 0 & 0 & 0 & 0 & 0 & \frac{-v_L \delta_0^L \delta_0^R}{\sqrt{2}} & \frac{- v_R \delta_0^L\delta_0^R }{2}& \frac{- \sqrt{3} v_R \delta_0^L \delta_0^R }{2}& 0 & 0 & \frac{v_L \delta_{1}^L \delta_0^R}{\sqrt{2}} \\
 0 & 0 & \frac{v_R \delta_0^L \delta_{-1}^R}{\sqrt{2}} & 0 & \frac{v_R \delta_0^L \delta_0^R}{\sqrt{2}} & 0 & 0 & \frac{-v_L \delta_0^L \delta_0^R}{\sqrt{2}} & 0 & 0 & 0 & 0 & \frac{v_L \delta_{1}^L \delta_0^R}{\sqrt{2}} \\ 
 0 &\frac{- v_R \delta_0^L \delta_{-1}^R }{2}& v_L \delta_{-1}^L \delta_0^R & \frac{- v_R \delta_0^L \delta_0^R }{2}& v_L \delta_0^L \delta_0^R & 0 & v_L \delta_0^L \delta_0^R & \frac{- v_R \delta_0^L \delta_0^R }{2}& 0 & 0 & 0 & -v_L \delta_{1}^L \delta_0^R & \frac{ v_R \delta_0^L \delta_1^R}{2} & 0 \\
 0 & \frac{- \sqrt{3} v_R \delta_0^L \delta_{-1}^R }{2}& 0 & \frac{- \sqrt{3} v_R \delta_0^L \delta_0^R }{2}& 0 & 0 & 0 & \frac{- \sqrt{3} v_R \delta_0^L \delta_0^R }{2}& 0 & 0 & 0 & 0 & \frac{\sqrt{3} v_R \delta_0^L \delta_1^R}{2} & 0 \\
 0 & 0 & 0 & 0 & \frac{v_R \delta_0^L \delta_{-1}^R}{\sqrt{2}} & 0 & 0 & 0 & -v_L \delta_{-1}^L\delta_0^R & 0 & 0 & 0 & \frac{v_R \delta_0^L \delta_0^R}{\sqrt{2}} \\
 0 & 0 & 0 & 0 & 0 & 0 & 0 & 0 & \frac{v_L \delta_{-1}^L \delta_0^R}{\sqrt{2}} & \frac{1}{2} v_R \delta_0^L \delta_{-1}^R & \frac{1}{2} \sqrt{3} v_R \delta_0^L \delta_{-1}^R & 0 & 0 & \frac{v_L \delta_0^L \delta_0^R}{\sqrt{2}} \\
 0 & 0 & 0 & 0 & 0 & 0 & \frac{v_R \delta_0^L \delta_{-1}^R}{\sqrt{2}} & \frac{v_L \delta_{-1}^L \delta_0^R}{\sqrt{2}} & 0 & 0 & 0 & \frac{v_R \delta_0^L\delta_0^R}{\sqrt{2}} & \frac{v_L \delta_0^L \delta_0^R}{\sqrt{2}} & 0
    \end{bmatrix}
    $}
    \label{eq:singlet_matrix}
\end{equation}

%\pagebreak

\begin{equation}
\rotatebox{90}{$
\begin{bmatrix}
    0 & 0 & 0 & 0 & -v_L \delta_1^L \delta_0^R & -v_R \delta_0^L \delta_1^R & 0 & 0 & 0 & 0 & 0 & 0 & 0 & 0 \\
   0 & 0 & 0 & 0 & 0 & -v_R \delta_0^L \delta_0^R & 0 & 0 & 0 & 0 & v_L \delta_1^L \delta_0^R & 0 & 0 & 0 \\
   0 & 0 & 0 & 0 & -v_L \delta_0^L \delta_0^R & 0 & 0 & 0 & 0 & 0 & 0 & v_R \delta_0^L \delta_1^R & 0 & 0 \\
   0 & 0 & 0 & 0 & -v_L \delta_0^L \delta_0^R & -v_R \delta_0^L \delta_0^R & 0 & 0 & 0 & 0 & 0 & 0 & 0 & 0 \\
   -v_L \delta_{-1}^L \delta_0^R & 0 & -v_L \delta_0^L \delta_0^R & -v_L \delta_0^L \delta_0^R & 0 & 0 & \frac{v_R \delta_0^L\delta_0^R}{\sqrt{2}} & 0 & 0 & v_L \delta_1^L \delta_0^R & 0 & 0 & \frac{v_R \delta_0^L\delta_1^R}{\sqrt{2}} & 0 \\
   -v_R \delta_0^L \delta_{-1}^R & -v_R \delta_0^L \delta_0^R & 0 & -v_R \delta_0^L \delta_0^R & 0 & 0 & \frac{v_L \delta_0^L \delta_0^R}{\sqrt{2}} & 0 & v_R\delta_0^L \delta_1^R & 0 & 0 & 0 & \frac{v_L \delta_1^L \delta_0^R}{\sqrt{2}} & 0 \\
   0 & 0 & 0 & 0 & \frac{v_R \delta_0^L \delta_0^R}{\sqrt{2}} & \frac{v_L \delta_0^L \delta_0^R}{\sqrt{2}} & 0 & 0 & 0 & 0 & \frac{v_R \delta_0^L \delta_1^R}{\sqrt{2}} & \frac{v_L \delta_1^L \delta_0^R}{\sqrt{2}} & 0 & 0 \\
   0 & 0 & 0 & 0 & 0 & 0 & 0 & 0 & 0 & 0 & v_L \delta_0^L \delta_0^R & v_R \delta_0^L \delta_0^R & 0 & 0 \\
   0 & 0 & 0 & 0 & 0 & v_R \delta_0^L \delta_{-1}^R & 0 & 0 & 0 & 0 & v_L \delta_0^L \delta_0^R & 0 & 0 & 0 \\
   0 & 0 & 0 & 0 & v_L \delta_{-1}^L \delta_0^R & 0 & 0 & 0 & 0 & 0 & 0 & v_R \delta_0^L \delta_0^R & 0 & 0 \\
   0 & v_L \delta_{-1}^L \delta_0^R & 0 & 0 & 0 & 0 & \frac{v_R \delta_0^L\delta_{-1}^R}{\sqrt{2}} & v_L \delta_0^L \delta_0^R & v_L \delta_0^L \delta_0^R & 0 & 0 & 0 & \frac{-v_R \delta_0^L\delta_0^R}{\sqrt{2}} & -v_L \delta_1^L \delta_0^R \\
   0 & 0 & v_R \delta_0^L \delta_{-1}^R & 0 & 0 & 0 & \frac{v_L \delta_{-1}^L \delta_0^R}{\sqrt{2}} & v_R \delta_0^L \delta_0^R & 0 & v_R \delta_0^L \delta_0^R & 0 & 0 & \frac{-v_L \delta_0^L \delta_0^R}{\sqrt{2}} & -v_R \delta_0^L \delta_1^R \\
   0 & 0 & 0 & 0 & \frac{v_R \delta_0^L \delta_{-1}^R}{\sqrt{2}} & \frac{v_L \delta_{-1}^L \delta_0^R}{\sqrt{2}} & 0 & 0 & 0 & 0 & \frac{-v_R \delta_0^L \delta_0^R}{\sqrt{2}} & \frac{-v_L \delta_0^L \delta_0^R}{\sqrt{2}} & 0 & 0 \\
   0 & 0 & 0 & 0 & 0 & 0 & 0 & 0 & 0 & 0 & -v_L \delta_{-1}^L \delta_0^R & -v_R \delta_0^L \delta_{-1}^R & 0 & 0
    \end{bmatrix}
    $}
\label{eq:doublet_matrix}
\end{equation}

\section{Zero-bandwidth-approximation calculation of spin-flip transition matrix elements}
\label{appB}

The transition matrix elements for pure spin-flip transitions can be obtained using a simplified zero-bandwidth approximation (ZBA) for the superconductors, neglecting all dynamical effects (order-parameter fluctuations). In this appendix we present the results of such a calculation and comment on the differences that originate from coupling to the transmon degrees of freedom in the full problem.

The Hamiltonian is $H = H_\mathrm{SC}^{(L)} + H_\mathrm{SC}^{(R)} + H_\mathrm{QD} + H_\mathrm{hop} + H_\mathrm{LR}$, with
\newcommand{\Vbeta}{V_\beta}
\newcommand{\VL}{V_L}
\newcommand{\VR}{V_R}
\begin{equation*}
\begin{split}
    H_\mathrm{SC}^{(\beta)} &= \sum_{k\sigma} \epsilon_k c^\dagger_{\beta k\sigma}c_{\beta k \sigma} + \Delta \sum_k e^{i \phi_\beta} c^\dagger_{\beta k \downarrow}c^\dagger_{\beta k\uparrow} + \text{H.c.}, \\
    H_\mathrm{QD} &= \epsilon n_{\mathrm{QD}} + U n_{\mathrm{QD}\uparrow}n_{\mathrm{QD}\downarrow} + E_z/2 (n_{\mathrm{QD}\uparrow}-n_{\mathrm{QD}\downarrow}), \\
    H_\mathrm{hop} &= \frac{v}{\sqrt{N}} \sum_{\beta = L,R} \sum_{k\sigma} d^\dagger_\sigma c_{\beta k\sigma} + \mathrm{H.c.} \\ 
    &+ \frac{i v_{\uparrow\downarrow}}{\sqrt{N}} \sum_{k\sigma} 
    \left( d^\dagger_\sigma c_{L, k\bar{\sigma}} + 
    c^\dag_{R, k\bar{\sigma}} d_{\sigma} + \mathrm{H.c.} \right), \\
    H_\mathrm{LR} &= \frac{t_\mathrm{sc}}{N} \sum_{k,k',\sigma}
      c^\dag_{L,k\sigma} c_{R,k'\sigma} + \text{H.c.} 
\end{split}
\end{equation*}
 We set $\phi_L=-\phi/2$ and $\phi_R=\phi/2$. We apply the gauge transformation $c_{\beta k \sigma} \to e^{i\phi_\beta/2} c_{\beta k\sigma}$ \cite{Oguri2004,Choi2004} to remove the phase from the pairing terms and transfer it to the hybridisation part.
We introduce the combinations of states $f_{\beta\sigma}=(1/\sqrt{N})\sum_k c_{\beta k \sigma}$, and drop all other modes from consideration. This leads to
\begin{equation*}
\begin{split}
    H_\mathrm{SC}^{(\beta)} &= \Delta f^\dagger_{\beta \downarrow} f^\dagger_{\beta \uparrow} + \text{H.c.}, \\
    H_\mathrm{hop} &= \sum_\sigma \Bigl[ \sum_{\beta = L,R} v e^{i\phi_\beta} d^\dagger_\sigma f_{\beta \sigma} + \text{H.c.} \\
    &+ i v_{\uparrow\downarrow} \left( e^{i\phi_L} d^\dag_\sigma f_{L\bar{\sigma}} + e^{-i\phi_R} f^\dag_{R\sigma} d_{\bar{\sigma}} + \text{H.c.} \right) \Bigr], \\
    H_\mathrm{LR} &= t_\mathrm{sc} \sum_\sigma e^{-i\phi_L+i\phi_R} f^\dag_{L\sigma}f_{R\sigma} + \text{H.c.}
\end{split}
\end{equation*}
Finally, we perform the Bogoliubov transformation on both superconductors:
\begin{equation}
\begin{split}
f_{\beta,\uparrow} &= (b_{\beta,\uparrow} + b^\dag_{\beta,\downarrow})/\sqrt{2},\\
f^\dag_{\beta,\downarrow} &= (b^\dag_{\beta,\downarrow} - b_{\beta,\uparrow})/\sqrt{2},
\end{split}
\end{equation}
and
\begin{equation}
\begin{split}
f_{\beta,\downarrow} &= (b_{\beta,\downarrow} - b^\dag_{\beta,\uparrow})/\sqrt{2}, \\
f^\dag_{\beta,\uparrow} &= (b^\dag_{\beta,\uparrow} + b_{\beta,\downarrow})/\sqrt{2}.
\end{split}
\end{equation}
In this basis the Hamiltonian is diagonal dominant: the only out-of-diagonal matrix
elements are couplings ($v$, $v_{\uparrow\downarrow}$, $t_\mathrm{sc}$).

For a decoupled system ($v=v_{\uparrow\downarrow}=t_\mathrm{sc}=0$), the low-energy subsector is spanned by the states $\ket{\sigma}=d^\dag_\sigma \ket{\mathrm{BCS}}$, where $\ket{\mathrm{BCS}}$ is the BCS ground states of both superconductors. Excited states are generated by creating quasiparticles in superconductors or adding/removing an electron on the QD site. There are in total 30 states in the $S_z=\pm 1/2$ sector of the Hilbert space, two in the low-energy subsector and 28 in the high-energy subsector. To study the effect of admixing high-energy states into the ground state doublet, we make use of the recursive Schrieffer-Wolff transformation (RSWT) \cite{Li2022}. The results in the lowest order are sufficiently compact to reproduce them here.

\begin{widetext}
The transition matrix element for the charge operator is
\begin{equation}
\langle \sigma | n | \bar{\sigma} \rangle
=
8 E_z (v^2+v_\mathrm{\uparrow\downarrow}^2)
\left( \frac{1}{(U+2\delta+2\Delta)^3} - \frac{1}{(U-2\delta+2\Delta)^3} \right)
\approx 
-\frac{96 E_z \delta (v^2+v_\mathrm{\uparrow\downarrow}^2)}{(U+2\Delta)^4}.
\end{equation}

The transition matrix element for the current operator $J=\mathrm{d}H/\mathrm{d}\phi$ is
\begin{equation}
\langle \sigma | J | \bar{\sigma} \rangle
=
2 E_z \delta t_\mathrm{sc} (v^2-v_{\uparrow\downarrow}^2) 
\frac{U^3 - 4 U \delta^2 + 18 U^2 \Delta + 8 \delta^2 \Delta + 60 U \Delta^2 + 56 \Delta^3}{\Delta (U-2\delta+2\Delta)^3 (U+2\delta+2\Delta)^3} \sin\phi
\approx 
\frac{2 E_z \delta t_\mathrm{sc} (v^2-v_{\uparrow\downarrow}^2) (U+14\Delta)}
{\Delta (U+2\Delta)^4}\sin\phi.
\end{equation}
(We note the difference in definition of current operator: here it is defined as the derivative with respect to the phase difference, for the full model it is defined as the derivative with respect to the external magnetic flux. At qualitative level these definitions should be equivalent, but there could be subtle differences in detailed behavior.)
\end{widetext}
As in the full model, a perpendicular field is needed to enable the spin-flip transitions. Both matrix elements go through zero at the particle-hole symmetric point $\delta=0$; in the full model, the zeros are shifted to some transmon-level-dependent non-zero value of $\delta$, see Figs.~\ref{fig:matel4} and \ref{fig:matel7}. The current matrix element also goes to zero for a given ratio of spin-flip and non-spin-flip tunneling rates, in the ZBA limit for $v_{\uparrow\downarrow}/v=1$, and in the full model for some renormalized values, see Fig.~\ref{fig:matel6}. Such cancellation does not occur for the charge matrix element, due to a different relative sign for the $v^2$ and $v_{\uparrow\downarrow}^2$ contributions. At this order of the approximation, the charge matrix elements do not depend on $\phi$, while the $\phi$-dependence of current matrix elements is sinusoidal. Neither of these results agrees with the behavior in the full model, which shows anharmonic $\phi$-dependence for $n$ and a phase shift of $\pi/2$ ($\cos\phi$ term) for $J$. This appears to be a limitation of the lowest order RSWT, not of the simplified ZBA description of the problem. At higher orders (or by performing an exact diagonalisation of the ZBA Hamiltonian) we find that both matrix elements are $\phi$-dependent (sinusoidal) with the same phase offsets as in the full problem (charge matrix element maximal at $\pi/2, 3\pi/2$ and passing through zero at $0$ and $\pi$, current matrix elements maximal at $0$ and $\pi$ and passing through zero at $\pi/2, 3\pi/2$). Furthermore, we find that zeros are shifted slightly away from the particle-hole symmetric point $\delta=0$, which is expected given that the ZBA Hamiltonian in the presence of SOC has no particular symmetry at $\delta=0$.

\section{Transmon transition matrix elements}
\label{appC}

\begin{figure}
    \centering
    \includegraphics[width=0.8\columnwidth]{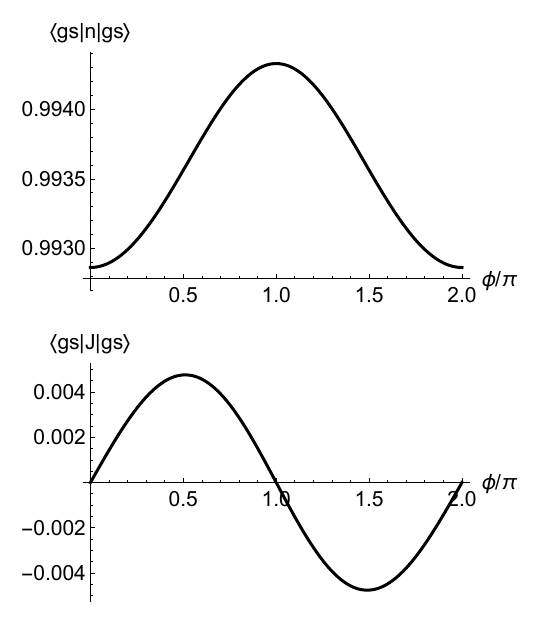}
    \caption{
 Ground state expectation value of charge and Josephson current operators in the zero-bandwidth-approximation model of the JJ. Parameters are: $U/\Delta=3$, $\epsilon/\Delta=-1$, $v/\Delta=0.2$, $v_{\uparrow\downarrow}/\Delta = 0.1$, $t_\mathrm{sc}/\Delta=0.1$, $E_z/\Delta=0.1$.
    }
    \label{fig:nJ}
\end{figure}

The model proposed in the previous section can also provide some clues about the $\phi$-dependence of the matrix elements for pure transmon transitions. For example, it is clear that the QD occupancy (charge) affects the value of the Josephson energy $E_J$, hence a modulation of dot filling can produce transitions between the different transmon levels. The coupling between the charge and SC phase degrees of freedom is quantified by the $\phi$-dependence of the diagonal matrix elements (expectation values) of the charge operator for the QD eigenstates. Observing that the $\phi$-dependence obtained in low-order RSWT is not reliable, we investigated this problem numerically using exact diagonalisation. We observe that the QD charge has a contribution to the expectation value that is proportional to $\cos\phi$, while the Josephson current has the expected contribution proportional to $\sin\phi$, see Fig.~\ref{fig:nJ}. Based on this, one would expect that the transmon transitions would be strongest for $\phi$ of strongest variation: close to $\pi/2$ for the charge operator and close to $0,\pi$ for the current operator. The first prediction is confirmed in the full model that has a sinusoidal $\phiext$ dependence, see Fig.~\ref{fig:matel1}a), but not the second: Fig.~\ref{fig:matel1}b) shows that the matrix element for $J$ actually remains fairly constant for all $\phiext$. 

\clearpage

\bibliography{bibl}

\end{document}